\documentclass[aps,superscriptaddress,nofootinbib,showpacs,eqsecnum,preprint,tightenlines]{revtex4}
\usepackage{hyperref}
\usepackage{epsfig,rotating}
\usepackage{amsmath,amssymb}
\usepackage{dsfont}

\newcommand{\be}{\begin{equation}}
\newcommand{\ee}{\end{equation}}
\newcommand{\bea}{\begin{eqnarray}}
\newcommand{\eea}{\end{eqnarray}}

\newcommand{\vx}{\vec{x}}

\newcommand{\vq}{\vec{q}}

\newcommand{\vk}{\vec{k}}

\newcommand{\nubar}{\overline{\nu}}

\begin{document}

\title{Condensates and quasiparticles in inflationary cosmology:\\  mass generation and decay widths.}
\author{Daniel Boyanovsky}
\email{boyan@pitt.edu} \affiliation{Department of Physics and
Astronomy, University of Pittsburgh, Pittsburgh, PA 15260, USA}

\date{\today}

\begin{abstract}
During   de Sitter inflation   massless particles of minimally coupled scalar fields  acquire a    mass and a decay width thereby  becoming \emph{quasiparticles}. For bare massless particles non-perturbative infrared radiative corrections lead to a self-consistent generation of mass, for a quartic self interaction  $M \propto \lambda^\frac{1}{4} H$, and for a cubic self-interaction  the mass is induced by the formation of  a non-perturbative \emph{condensate}   leading to    $M \propto \lambda^\frac{1}{3} H^\frac{2}{3}$ .   These radiatively generated masses restore de Sitter invariance and result in anomalous scaling dimensions of superhorizon fluctuations.   We introduce a generalization  of the non-perturbative Wigner-Weisskopf method to obtain the time evolution of quantum states that include the self-consistent generation of mass and regulate the infrared behavior.    The infrared divergences  are manifest as poles in $\Delta=M^2/3H^2$ in the single particle self-energies, leading to a re-arrangement of the perturbative series    non-analytic in the couplings. A set of simple rules that yield the leading order infrared contributions to the decay width are obtained and implemented.
 The lack of kinematic thresholds entail that all particle states acquire a decay width, dominated by the emission and absorption of   superhorizon quanta $\propto (\lambda/H)^{4/3}\,[H/k_{ph}(\eta)]^6~;~\lambda\,[H/k_{ph}(\eta)]^6 $ for cubic and quartic couplings respectively to leading order in $M/H$.   The decay of single particle quantum states hastens as their wavevectors cross the Hubble radius and their width is related to the highly squeezed limit of the bi- or tri-spectrum of scalar fluctuations respectively.
\end{abstract}

\pacs{98.80.-k,98.80.Cq,11.10.-z}

\maketitle

\section{Introduction}\label{sec:intro}
Quantum fluctuations during inflation seed the inhomogeneities which  are manifest as   anisotropies in the cosmic microwave background and are responsible for large scale structure formation. In its simplest inception the inflationary stage can be effectively described as a quasi-deSitter space time.  Early studies\cite{polyakov1,IR1,IR2,allen,folaci,dolgov} revealed that de Sitter space time  features infrared instabilities and profuse particle production in interacting field theories. Infrared divergences in loop corrections to correlation functions hinder the reliability of the perturbative expansion\cite{weinberg,seery,branrecent},  led to the suggestion of an infrared instability of the vacuum\cite{polyakov,kroto,akhmedov,higuchi,vidal}, and affect  correlation functions during inflation\cite{weinberg,giddins,seery,bran,mazumdar,leblond2} requiring a non-perturbative treatment.

 Cosmological expansion modifies the energy-uncertainty relation allowing ``virtual'' excitations to persist longer, leading to remarkable phenomena, which is stronger in de Sitter space  time as clarified in  ref.\cite{woodard}.
 Particle production in a de Sitter background has been argued to provide a  dynamical``screening'' mechanism that leads to relaxation of the cosmological constant\cite{emil,IR3,branmore} through back reaction,  much like the production of particle-antiparticle pairs in a constant electric field. The possibility that   back reaction from the production of virtual excitations may yield  a dynamical mechanism of evolution of dark energy  rekindled the interest on infrared effects in de Sitter space time. A body of work established that infrared and secular divergences are manifest in super-Hubble fluctuations during de Sitter (or nearly de Sitter) inflation\cite{petri,enq,riotto} and also invalidate the semiclassical approximation\cite{holman}, thus a  consistent  program that provides a resummation of the perturbative expansion is required. One possible approach is furnished by the dynamical renormalization group\cite{drg} which provides a non-perturbative resummation of the secular divergences and has been implemented in several studies in de Sitter space time\cite{boyan}   and   suggests a dynamical generation of mass\cite{holman}. The generation of a mass through the build up of infrared fluctuations was originally anticipated in the seminal work of ref.\cite{staroyoko}, and explored and extended in ref.\cite{richard}, and more recently  a self-consistent mechanism of  mass generation for scalar fields through infrared fluctuations has been suggested\cite{petri,holman,rigo,garb,arai,serreau}.

 Another particular aspect of the rapid cosmological expansion is the lack of a global time-like killing vector which leads to remarkable physical effects in de Sitter space time, as it implies the lack of particle thresholds (a direct consequence of energy-momentum conservation) and the decay of   fields even in their own quanta\cite{boyprem,boyan} with the concomitant particle production, a result that was  confirmed in ref.\cite{moschella,akhmedov} and more recently investigated in ref.\cite{donmor,leblond} for the case of heavy fields.
 For light scalar fields in de Sitter space time with   mass   $M \ll H$, it was shown in ref.\cite{boyan}  that the infrared enhancement of self-energy corrections is manifest as   poles in $\Delta = M^2/3 H^2$ and that the most infrared singular contributions to the self-energy can be isolated systematically in an expansion in $\Delta$ akin to the $\epsilon$ expansion in critical phenomena. A similar expansion was noticed in refs.\cite{holman,leblond,rigo,smit}.

Most of the efforts towards understanding infrared effects in de Sitter (or quasi de Sitter) cosmology focus  on correlation functions, and only recently the issue of the time evolution of the \emph{quantum states} has began to be addressed. In ref.\cite{boyhol} the Wigner-Weisskopf method\cite{ww,boyaww} ubiquitous in quantum optics\cite{qoptics} has been adapted and extended as a non-perturbative quantum field theory method in inflationary cosmology which allows to study the time evolution of quantum states. This method   reveals how quantum states \emph{decay} in time and it has been shown to be  equivalent to the dynamical renormalization group in Minkowski space time\cite{boyhol}.

 \vspace{2mm}

\textbf{Motivation and results:}

There are at least two reasons to study the evolution of \emph{quantum states} by implementing a non-perturbative resummation method: i) questions of instability of vacuum or excited \emph{states} cannot be directly addressed by studying correlation functions perturbatively, ii) infrared and secular divergences in correlation functions require a resummation of the perturbative series. Whereas the dynamical renormalization group\cite{drg,holman,leblond} provides a resummation scheme in some cases, alternative non-perturbative resummation methods may offer novel insights and  undoubtedly will be a welcome addition to the non-perturbative tools to study dynamical phenomena in cosmology.

 The main observation is that in the interaction picture  field operators feature the free field time dependence and all the interaction effects are contained in the time evolution of states, therefore a method that provides a non-perturbative resummation scheme for the time evolution of states \emph{may} provide an equivalent resummation framework for correlation functions by saturating the intermediate states with the time evolved states obtained from the non-perturbative time evolution.

 In this article we combine the   expansion in $\Delta$ advocated in ref.\cite{boyan} with the Wigner-Weisskopf method introduced in ref.\cite{boyhol} to study the nature of the \emph{single particle} excitations during de Sitter inflation.

We focus our study on \emph{massless minimally coupled} scalar field theories  with typical cubic or quartic interactions   and find that, similarly to finite temperature field theory, these excitations become \emph{quasiparticles} acquiring a \emph{self-consistent} mass that regulates the infrared as found in refs.\cite{staroyoko,holman,leblond} but also \emph{a decay width}.

In section II we study the generation of \emph{radiatively induced mass} through the build up of infrared effects in a self-consistent manner. In the case of cubic self-interaction vertex the strong infrared behavior leads to the formation of a \emph{condensate} that reveals that the theory is driven to a new stable minimum by radiative corrections, the expectation value of the field in this state yields a self-consistently induced mass. In the case of a quartic self-interaction the resummation of tadpole-type diagrams which are infrared
divergent in the massless theory lead to the self-consistent generation of a mass,    confirming the results of refs.\cite{staroyoko,holman}. This self-consistent  mechanism regulates the infrared behavior and induces, radiative anomalous scaling dimensions of superhorizon fluctuations. The infrared singularities for massless particles are replaced by poles in $\Delta = M^2/3H^2$ with $M$ being the radiatively generated mass.

In section III we combine the self-consistent approach with a generalization of the non-perturbative Wigner-Weisskopf method to obtain the long time evolution of single particle \emph{quantum states}.

We find that single particle states decay via the emission and absorption of superhorizon quanta and obtain their decay ``widths'' both for super and sub horizon modes to leading order in the expansion in $\Delta$. The self-consistent mass generation regulates the infrared behavior which is now manifest as poles in $\Delta$ and leads to a rearrangement of the perturbative expansion which is \emph{non-analytic} in the couplings. The decay of quantum states hastens as their wavevectors cross the Hubble radius. We argue that the order of the poles in $\Delta$ reflect the number of superhorizon quanta emitted in the decay process and obtain a set of simple rules to extract the leading order contributions in $\Delta$ to the decay ``widths''. We provide an intepretation of the decay ``width'' of superhorizon modes in terms of a relation between the single particle self energy and the bispectrum (for cubic coupling) or tri-spectrum (for quartic coupling) of scalar fluctuations in a highly squeezed limit. The order of the pole in $\Delta$ reflects the number of squeezed sides in the bi-or tri-spectrum configuration respectively.

Conclusions, comments and further questions are presented in section IV. An appendix is devoted to the calculation of the self-energy for a cubic coupling to leading and next to leading order in $\Delta$.

\section{\label{sec:mass} Condensate and self-consistent mass generation:}

We consider scalar field theories in a spatially flat Friedmann-Robertson-Walker (FRW)
cosmological spacetime with scale factor $a(t)$. In comoving
coordinates, the action is given by
\begin{equation}
S= \int d^3x \; dt \;  a^3(t) \Bigg\{
\frac{1}{2}{\dot{\phi}^2}-\frac{(\nabla
\phi)^2}{2a^2}-\frac{1}{2}\Big(M^2+\xi \; \mathcal{R}\Big)\phi^2
- \lambda \;  \phi^{\,p}  \Bigg\},\quad p=3,4 \label{lagrads}
\end{equation}
with \be \mathcal{R} = 6 \left(
\frac{\ddot{a}}{a}+\frac{\dot{a}^2}{a^2}\right) \ee being the Ricci
scalar,   $\xi=
0,1/6$ correspond  to minimal coupling  and
conformal coupling respectively.

Specializing now to the de Sitter case with $a(t) = e^{H t}$, it is convenient to pass to conformal time $\eta = -e^{-Ht}/H$ with $d\eta =
dt/a(t)$ and introduce a conformal rescaling of the fields
\begin{equation}
a(t)\phi(\vx,t) = \chi(\vx,\eta).\label{rescale}
\end{equation}
The action becomes (after discarding surface terms that will not
change the equations of motion) \be S  =
  \int d^3x \; d\eta  \; \Bigg\{\frac12\left[
{\chi'}^2-(\nabla \chi)^2-\mathcal{M}^2 (\eta) \; \chi^2  \right] -\lambda \big[C (\eta)\big]^{(4-p)} \;  \chi^{\,p}   \Bigg\} \; , \label{rescalagds}\ee
with primes denoting derivatives with respect to
conformal time $\eta$ and \be \mathcal{M}^2 (\eta) =
\Big(M^2 +\xi \mathcal{R}\Big)
C^2(\eta)-\frac{C''(\eta)}{C(\eta)}  \; , \label{massds}
\ee
where for de Sitter spacetime \be C(\eta)= a(t(\eta))= -\frac{1}{H\eta}. \label{scalefactords}\ee  In this case the effective time dependent mass is given by
\be
\mathcal{M}^2 (\eta)  = \Big[\frac{M^2 }{H^2}+12\Big(\xi -
\frac{1}{6}\Big)\Big]\frac{1}{\eta^2}   \; , \label{massds2}
\ee
\noindent in what follows we consider the case of minimal coupling to gravity, namely $\xi =0$.
The Heisenberg equations of motion for the spatial Fourier modes
of wavevector $k$  of the fields in the non-interacting ($\lambda=0$)
theory are given by
\be  \chi''_{\vk}(\eta)+
\Big[k^2-\frac{1}{\eta^2}\Big(\nu^2 -\frac{1}{4} \Big)
\Big]\chi_{\vk}(\eta)  =   0   \label{dsmodes}  \; ,
\ee
\noindent where
\be
\nu^2   =  \frac{9}{4}- \frac{M^2 }{H^2}
   \label{nu}\ee
We will choose Bunch-Davies vacuum conditions for which the two linearly independent solutions are given by
\bea
g_{\nu}(k;\eta) & = & \frac{1}{2}\; i^{\nu+\frac{1}{2}}
\sqrt{-\pi \eta}\,H^{(1)}_\nu(-k\eta)\label{gnu}\\
f_{\nu}(k;\eta) & = & \frac{1}{2}\; i^{-\nu-\frac{1}{2}}
\sqrt{-\pi \eta}\,H^{(2)}_\nu(-k\eta)= g^*_{\nu}(k;\eta) \label{fnu}  \; ,
\eea
 \noindent where $H^{(1,2)}_\nu(z)$ are Hankel functions. Expanding the field operator in this basis in a comoving volume $V$
\be \chi(\vec{x},\eta) = \frac{1}{\sqrt{V}}\sum_{\vec{k}} \Big[a_{\vec{k}}\,g_\nu(k;\eta)\,e^{i\vec{k}\cdot\vec{x}}+ a^\dagger_{\vec{k}}\,\,g^*_\nu(k;\eta)\,e^{-i\vec{k}\cdot\vec{x}}\Big]\,. \label{quantumfieldds1}\ee
The Bunch-Davies vacuum is defined so that \be a_{\vec{k}}|0\rangle = 0 \,,\label{dsvac}\ee and the Fock  states are obtained   by applying creation operators $a_{\vec{k}}^{\dagger}$ to the vacuum.

 In the Schroedinger picture the quantum states $|\Psi(\eta)\rangle$ obey
 \be i\frac{d}{d\eta}|\Psi(\eta)\rangle = H(\eta) \, |\Psi(\eta)\rangle \label{Spic}\ee where in an expanding cosmology the Hamiltonian $H(\eta)$ is generally a function of $\eta$. Introducing the time evolution operator $U(\eta,\eta_0)$ obeying
 \be i\frac{d}{d\eta} U(\eta,\eta_0) = H(\eta)\,U(\eta,\eta_0) , \quad U(\eta_0,\eta_0) = 1, \label{Uds}\ee the solution of the Schroedinger equation is $|\Psi(\eta)\rangle = U(\eta,\eta_0)\,|\Psi(\eta_0)\rangle $. Writing the Hamiltonian as $H(\eta) = H_0(\eta) + H_{i}(\eta)$ with $H_0(\eta)$ the non-interacting Hamiltonian, and introducing the time evolution operator of the free theory $U_0(\eta,\eta_0)$ satisfying
 \be i\frac{d}{d\eta} U_0(\eta,\eta_0) = H_0(\eta)\, U_0(\eta,\eta_0), \quad i\frac{d}{d\eta} U^{-1}_0(\eta,\eta_0) = - U^{-1}_0(\eta,\eta_0)\, H_0(\eta) , \quad U_0(\eta_0,\eta_0) =1, \label{U0ds}\ee the interaction picture states are defined as
 \be |\Psi(\eta)\rangle_I = U_I(\eta,\eta_0)|\Psi(\eta_0)\rangle_I =  U^{-1}_0(\eta,\eta_0) |\Psi(\eta)\rangle. \label{ipds}\ee where $U_I(\eta,\eta_0)$ is the time evolution operator in the interaction picture obeying
 \be  \frac{d}{d\eta}U_I(\eta,\eta_0) = -i H_I(\eta) U_I(\eta,\eta_0), \quad U_I(\eta_0,\eta_0)=1 \label{UI}\ee  where the  interaction Hamiltonian in the interaction picture \be H_I(\eta) = U^{-1}_0(\eta,\eta_0) H_{i}(\eta) U_0(\eta,\eta_0) \label{HIdsdef}\ee     is given by
 \be H_I(\eta) = \frac{\lambda}{[-H\eta]^{(4-p)}}\int  ~\big[\chi(\vec{x},\eta)\big]^p~ d^3 x \label{HIds}\ee and $\chi$ is the free field Heisenberg field operator in eq.(\ref{quantumfieldds1}).

 In perturbation theory
 \be U_I(\eta,\eta_0) = \Bigg[1-i \int^{\eta}_{\eta_0} d\eta' H_I(\eta') +\cdots\Bigg] \,.\label{timevol}\ee
 In the interaction picture operators evolve in time with the free Hamiltonian $H_0$ whereas states evolve as in eqn. (\ref{ipds}).

\subsection{The tadpole and the $\Delta$ expansion:} The tadpole will play an important role in the mechanism of self-consistent mass generation, it is given by
\be \langle 0| \chi^2(\vec{x},\eta)|0 \rangle = \int \frac{d^3k}{(2\pi)^3} \big|g_\nu(k,\eta)\big|^2 = \frac{1}{8\pi\,\eta^2}~\int \frac{dz}{z}~ z^3~\big|H^{(1)}_\nu(z)\big|^2 \label{tadpole}\ee for the massless, minimally coupled case $\nu = 3/2$ and
\be z^3~\big|H^{(1)}_\nu(z)\big|^2  = \frac{2}{\pi}\,[1+z^2] \label{tadmin}\ee in which case the integral features both the usual quadratic and logarithmic ultraviolet divergence, but also a logarithmic infrared divergence. For minimally coupled ``light'' fields with $M^2/H^2 \ll 1$ it follows that
\be \nu = \frac{3}{2}-\Delta ~~;~~ \Delta = \frac{M^2 }{3H^2} +\cdots \label{deltadef}\ee and
\be  z^3 \, \left|H^{(1)}_\nu(z)\right|^2 \buildrel{z \to
0}\over=\left[ \frac{2^{\nu} \; \Gamma(\nu)}{\pi} \right]^2 \; z^{2
\, \Delta} \label{smallz}\ee thus $\Delta > 0$ regulates the infrared behavior of the tadpole. To isolate the infrared we introduce ultraviolet ($\Lambda_p$) and infrared ($\mu_p$) cutoffs in physical momenta and write the integral (\ref{tadpole}) as \be\label{intsplit}
\int^{\frac{\Lambda_p}{H}}_0 \frac{dz}{z} \; z^3 \; |H^{(1)}_\nu(z)|^2=
\int^{\frac{\mu_p}{H}}_0 \frac{dz}{z} \; z^3 \, \left|H^{(1)}_\nu(z)\right|^2
+ \int^{\frac{\Lambda_p}{H}}_{\frac{\mu_p}{H}} \frac{dz}{z} \; z^3 \,
\left|H^{(1)}_\nu(z)\right|^2 \, . \ee \noindent $\mu_p \rightarrow 0 $ acts here
as  infrared cutoff for the first integral. The second integral is
ultraviolet and infrared finite for finite $\mu_p, \; \Lambda_p$. We
can set   $\nu=3/2$ in this integral and use   (\ref{tadmin}).
In the first integral we obtain the leading order contribution, namely the pole and leading logarithm, by
using the small argument limit of the Hankel functions (\ref{smallz}) and we find that eq.(\ref{intsplit}) yields after
calculation, \be\label{IRint} \int^{\frac{\mu_p}{H}}_0 \frac{dz}{z} \; z^3 \,
\left|H^{(1)}_\nu(z)\right|^2 = \frac{2}{\pi}\left[\frac{1}{2 \,
\Delta}+ \frac{ {\mu^2_p}}{2H^2} + \gamma - 2 + \ln   \frac{2 \mu_p}{H}
+\mathcal{O}(\Delta)\right]\,, \ee \noindent where we have displayed
the pole in $\Delta$ and the leading infrared logarithm. In the second integral in (\ref{intsplit}) we set $\nu =3/2$ and combining its result with (\ref{IRint} ) we find that  the dependence on the infrared cutoff $\mu_p$ cancels in the limit $\mu_p\rightarrow 0$ leading to the following   final result for the tadpole
\be\label{QC}
  \langle 0|\chi^2(\vx,\eta)|0\rangle = \frac{1}{8\pi^2 \,\eta^2}
\left[ \frac{{\Lambda_p}^2}{H^2} + 2 \ln \frac{\Lambda_p }{H}+\frac1{\Delta}
 + 2 \, \gamma - 4 + \mathcal{O}(\Delta) \right]\,,
\ee
\noindent
where $\gamma$ is the Euler-Mascheroni constant.  While the
quadratic and logarithmic \emph{ultraviolet} divergences are
regularization scheme dependent, the pole in $\Delta$ arises from
the  infrared behavior and is independent of the regularization
scheme. In particular this pole coincides with that found
in the expression for $<\phi^2(\vx,t)>$ in refs.\cite{boyan,holman,rigo,smit}. The
\emph{ultraviolet divergences}, in whichever renormalization scheme,
 require that the effective field theory be
defined to contain \emph{renormalization counterterms} in the bare
effective Lagrangian, for the tadpole this counterterm is of the form $\chi(\eta)\,J(\eta)$  and $J(n)$ is required to cancel    the ultraviolet divergences. Thus,   the { \it{renormalized}} tadpole
\be \mathcal{I}(\eta)\equiv \langle 0| \chi^2(\vx,\eta)|0 \rangle_{ren} = \frac{1}{8\pi^2 \,\eta^2}~\frac1{\Delta} ~\big[1+
  \cdots \big]\,, \label{poletad}
\ee
\noindent where the dots stand for higher order terms in $\Delta \ll 1$.

\subsection{Self-consistent mass generation:}

\subsubsection{$\lambda \chi^3 $ theory: condensate formation} 
\begin{figure}[h!]
\begin{center}
\includegraphics[height=1 in,width=1 in,keepaspectratio=true]{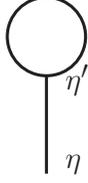}
\caption{Tadpole contribution to the expectation value $ \langle  \widetilde{0}\big|\chi(\vec{y},\eta) \big|\widetilde{0}\rangle$.}
\label{fig:tadpole}
\end{center}
\end{figure}

In this theory radiative corrections induce an expectation value of the field in the ``dressed'' vacuum state, which up to first order in perturbation theory is given by
\be \big|\widetilde{0}(\eta)\rangle = \Bigg[1+i \frac{\lambda}{H} \int^{\eta}_{\eta_0} \frac{d\eta'}{\eta'} \int d^3 \vx ~~ \chi^3(\vx,\eta') \Bigg]\,\big|0\rangle \,,\label{dresvac}\ee and the expectation value of $\chi$ to leading order in $\lambda$ is given by
\be \langle  \widetilde{0}\big|\chi(\vec{y},\eta) \big|\widetilde{0}\rangle = 3 i \frac{\lambda}{H} \int^{\eta}_{\eta_0} \frac{d\eta'}{\eta'} \int d^3 \vx ~~\Big[\chi(\vec{y},\eta), \chi(\vx,\eta')\Big] ~\langle 0|\chi^2(\vx,\eta')|0\rangle \,,\label{expval} \ee \noindent this expression is depicted in fig.(\ref{fig:tadpole}). We find
\be \langle  \widetilde{0}\big|\chi(\vec{y},\eta) \big|\widetilde{0}\rangle = -  \frac{3\,\lambda}{16\,\pi \,H\,\nu} \int^{\eta}_{\eta_0} \frac{d\eta'}{\eta'^{\,3}}\Big[(\eta)^{\beta_+}\,(\eta')^{\beta_-}- (\eta')^{\beta_+}\,(\eta)^{\beta_-}\Big]\, \int \frac{dz}{z}~ z^3~\big|H^{(1)}_\nu(z)\big|^2 \label{expval2} \ee where \be \beta_{\pm} = \frac{1}{2} \pm \nu \,. \label{betas} \ee

In order to understand the nature of the infrared divergences, let us first consider the massless case, namely $\Delta=0$, for which we find
\be \langle  \widetilde{0}\big|\chi(\vec{y},\eta) \big|\widetilde{0}\rangle = -  \frac{3\,\lambda}{16\,\pi \,H\,\nu} \,\frac{1}{\eta}\Bigg[\ln\Big(\frac{\eta_0}{\eta}\Big)-\frac{1}{3}\Bigg]\,\int \frac{dz}{z}~ z^3~\big|H^{(1)}_\nu(z)\big|^2 \,. \label{expvalm0} \ee There are \emph{two} sources of infrared singular physics in this expression, i) for $\nu = 3/2$ the $z$ integral is infrared divergent (it is also ultraviolet divergent but this divergence can be canceled by the counterterm  discussed above), ii) in the long time limit $  \eta \rightarrow 0^-$ the logarithmic term is \emph{secular} and entails that the (unscaled) expectation value grows in time: the factor $-1/H\eta = a(t)$ reflects that the expectation value of  the unscaled field $\phi = \chi/a(t)$ (\ref{rescale}) would be \emph{constant} were it not for the logarithmic term. The growth of the expectation value implies the formation of a \emph{condensate}. For the massless case $\langle  \widetilde{0}\big|\chi(\vec{y},\eta) \big|\widetilde{0}\rangle \simeq \ln^2[\eta/\eta_0]$ if the integral is regulated with an infrared cutoff  constant in comoving coordinates\cite{holman}.

For $M^2  \neq 0 $ \emph{both} infrared divergences are regulated, using the result (\ref{poletad}) and $\Delta \ll 1 $ we find for $\eta/\eta_0 \rightarrow 0$,
\be \langle  \widetilde{0}\big|\chi(\vec{y},\eta) \big|\widetilde{0}\rangle =   \overline{\chi}(\eta) \rightarrow  -  \frac{ \lambda}{8\,\pi^2 \,H\,\Delta^2\,\eta} \,\big[1+ \mathcal{O}(\Delta) +\cdots\big]\,.\label{chiex}\ee The un-scaled field $\phi$ acquires a \emph{constant} expectation value asymptotically for $\eta/\eta_0 \rightarrow 0$,
\be \langle  \widetilde{0}\big|\phi(\vec{y},\eta) \big|\widetilde{0}\rangle = \frac{1}{C(\eta)}~\langle  \widetilde{0}\big|\chi(\vec{y},\eta) \big|\widetilde{0}\rangle \rightarrow     \frac{ \lambda}{8\,\pi^2  \,\Delta^2 } \,\big[1+ \mathcal{O}(\Delta) +\cdots\big]\,.\label{phiex}\ee

\begin{figure}[h!]
\begin{center}
\includegraphics[height=1in,width=1in,keepaspectratio=true]{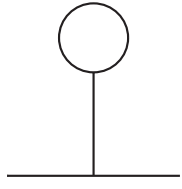}
\caption{The tadpole and expectation value contribution to the self-energy in $\lambda \, \chi^3$. }
\label{fig:fi3mass}
 \end{center}
\end{figure}

Since the field is acquiring an expectation value we shift the field and define
\be \chi(\vx,\eta) = \Psi(\vx,\eta) +  \overline{\chi}(\eta)~~;~~\langle  \widetilde{0}\big|\Psi(\vec{x},\eta) \big|\widetilde{0}\rangle =0\,, \label{shift} \ee \noindent introducing this shift in the interaction Hamiltonian (\ref{HIds}) for $p=3$ we find
\be H_I = \int d^3x \Bigg[ \frac{1}{\eta ^2} \, \frac{M^2 }{2\,H^2} \,\Psi^2 -\frac{\lambda}{H\eta} \Psi^3 \Bigg]\label{shiftedHI}\ee where
\be  \frac{1}{\eta ^2} \, \frac{M^2 }{2\,H^2} = -3\frac{\lambda}{H\eta}\,\overline{\chi}(\eta) \,, \label{mas3s}\ee leading to
\be \frac{M^2 }{H^2} = \frac{3\, \lambda^2}{4\,\pi^2 \, H^2 \,\Delta^2 }\,\Big[1+\mathcal{O}(\Delta)+\cdots\Big]\,, \label{masita}\ee and we neglected terms that are constant and linear in $\Psi$ (see the discussion below). This suggests a mechanism of \emph{self-consistent mass generation}, indeed interpreting $M $ as the mass of the field, with $\Delta$ given by  (\ref{deltadef}) eqn. (\ref{masita}) becomes a self-consistent condition with the solution
\be M  = H~\sqrt{3}\,\Big[\frac{\lambda}{2\pi H}\Big]^{1/3}\,. \label{selfmasa}\ee This mass term is identified with a self-energy contribution depicted in fig. (\ref{fig:fi3mass}).

A systematic implementation  of this self-consistent mechanism can be formulated as follows:

\begin{itemize}
\item As in  renormalized perturbation theory, a perturbative expansion around the vacuum state with the correct mass including radiative corrections is achieved by
adding a mass term to the non-interacting Lagrangian density and subtracting it in the interacting part as a \emph{mass counterterm}, namely
\bea \mathcal{L} & = & \mathcal{L}_0+\mathcal{L}_I \nonumber \\
\mathcal{L}_0  & = & \frac12\left[
{\chi'}^2-(\nabla \chi)^2-\frac{1}{\eta^2}\Big(\frac{M^2}{H^2}-2\Big)  \; \chi^2  \right] \label{L0} \\ \mathcal{L}_I  &  =  &  \frac{M^2}{2\,H^2\,\eta^2}   \; \chi^2 -\frac{\lambda}{[-H\eta] }\;\chi^3 \label{LI}
\eea

\item Expand the field in Fourier modes in a finite spatial comoving volume $V$ quantizing the modes with the usual periodic boundary conditions in this volume
\bea \chi(\vec{x},\eta) & = & \frac{1}{\sqrt{V}}\sum_{\vec{k}} \tilde{\chi}(\vk,\eta)\;e^{i\vec{k}\cdot\vec{x}}\label{FT}\\
\tilde{\chi}(\vk,\eta) & = & a_{\vec{k}}\,g_\nu(k;\eta) + a^\dagger_{-\vec{k}}\,\,g^*_\nu(k;\eta) \label{chitilk}
\eea where the mode functions are given by (\ref{gnu}) with $\nu$ given by (\ref{nu}).

\item
Separate the \emph{zero mode} $\tilde{\chi}(\vec{0},\eta)$ and normal order the full Hamiltonian in the interaction picture of the free \emph{massive} field with mass $M $. The cubic interaction becomes
\bea \frac{\lambda}{H\eta}\int d^3 x ~\chi^3(\vec{x},\eta) & = & \frac{\lambda}{\sqrt{V} H\eta }\Bigg\{3 \Big[    \sum_{\vk \neq 0}  : \tilde{\chi}(\vk,\eta)\,\tilde{\chi}(-\vk,\eta):\;+ {V}\,\mathcal{I}(\eta)\Big]\tilde{\chi}(\vec{0},\eta)\nonumber\\ &  &
+    \sum_{\vq\neq \vk\neq 0} :\tilde{\chi}(\vk,\eta)\tilde{\chi}(\vq,\eta)\,\tilde{\chi}(-\vk-\vq,\eta):+ \tilde{\chi}^3(\vec{0},\eta)
\Bigg\} \label{NOL}\eea where we have renormalized the tadpole by canceling the ultraviolet divergences with appropriate counterterms yielding the finite contribution $\mathcal{I}(\eta)$ given by (\ref{poletad}). In the $V\rightarrow \infty $ limit the expectation value of the field
\be \langle \chi(\vec{x},\eta) \rangle \equiv \overline{\chi}(\eta) = \frac{1}{\sqrt{V}} \langle \tilde{\chi}(\vec{0},\eta) \rangle \label{chitiex}\ee is now determined by the linear term $\sqrt{V}\,\mathcal{I}(\eta)\, \tilde{\chi}(\vec{0},\eta)$ and we find from the result  (\ref{chiex})
\be \langle \tilde{\chi}(\vec{0},\eta) \rangle = -\frac{\lambda\,\sqrt{V}}{8\pi^2\, H\,\eta\,\Delta^2 } \label{chitiex2} \ee

\item The presence of a vacuum expectation value of $\chi$ depicted in fig. (\ref{fig:tadpole}) indicates that an effective action for a long-wavelength component of the field features a $\emph{linear}$ term in $\chi$ along with the induced mass term from this expectation value indicated in fig. (\ref{fig:fi3mass}). Such a linear term indicates a new, radiatively induced minimum of the effective potential and requires that the field be shifted by the new vacuum expectation value, so that the one-point function (expectation value) of the shifted field vanishes in the new (correct) vacuum. Therefore, as usual, quantization around the condensed state is achieved by shifting the field by its expectation value,   defining $\Psi$ as in eqn. (\ref{shift}), carrying out the Fourier transform and introducing the linear counterterm $\mathcal{J}(\eta)$ that ensures the vanishing of the expectation value of $\Psi$ one finds\footnote{A linear term $\frac{M^2}{H^2\eta^2} \overline{\chi}\,\Psi$ is cancelled between the free and counterterm parts of the Lagrangian density.}
    \be  \mathcal{L}_I   =    \frac{1}{2\eta^2} \Big[\frac{M^2}{ H^2\, }-  \frac{3  \lambda^2}{4\,\pi^2 \, H^2 \,\Delta^2 }\Big]    \Psi^2  +\,\frac{\lambda}{ H\eta  } :\Psi^3: + ~  \Big \{\mathcal{J}(\eta) +  \frac{3\lambda}{ H\,\eta} \; \Big[\mathcal{I}(\eta)+ \overline{\chi}^{\,2}(\eta)\Big] \Big \}\Psi \label{shiftedLI}\ee  with $:\Psi^3: = \Psi^3 - 3\, \mathcal{I}(\eta)$ where
    $\mathcal{I}(\eta)$ is given by (\ref{poletad}) and we have neglected constant terms. The linear counterterm $\mathcal{J}(\eta)$ is fixed by requiring that  $\langle 0|\Psi |0\rangle =0$ in the correct vacuum state systematically in perturbation theory. To one-loop  order
    \be \mathcal{J}(\eta) +  \frac{3\lambda}{ H\,\eta} \; \Big[\mathcal{I}(\eta)+ \overline{\chi}^{\,2}(\eta)\Big] =0 \label{onelupcount}\,. \ee

      This choice of  counterterm simultaneously    ensures that $\langle 0|\Psi |0\rangle =0$ and the  cancellation of  the matrix element of the interaction between the vacuum and the single particle state with zero momentum $\langle 1_{\vec{0}}|H_I|0\rangle =0$, thus ensuring harmonic perturbations around the new vacuum state.
     Cancellation of the quadratic term in $\mathcal{L}_I$ yields the self-consistent \emph{gap} equation leading to the \emph{radiatively generated mass} given by eqn. (\ref{selfmasa}), however, we leave the mass counterterm in $\mathcal{L}_I$ anticipating the possibility of a further one loop contribution from the non-local self energy diagram studied in the next section.

      It is important to highlight that renormalizing the short distance divergence with different renormalization schemes in $<\Psi^2>$ \emph{does not affect} the leading order pole in $\Delta$, a purely infrared effect, as different subtraction schemes differ by finite constants. Thus the self-generated mass term is a genuine radiative infrared effect.
 \end{itemize}

The fact that one (zero momentum) mode of the $\chi$ field becomes \emph{macroscopically} occupied, with $\langle \tilde{\chi}(\vec{0},\eta) \rangle^2 \propto V$ in the $V\rightarrow \infty$ limit (see eqn. (\ref{chitiex2})) signals the emergence of a \emph{condensate} as a consequence of radiative corrections.  As mentioned above the $\eta$ dependence of this condensate entails that the zero momentum Fourier mode of the unscaled field $\phi$ acquires a \emph{constant  and macroscopically large} expectation value.     In Minkowski space time as well as in de Sitter space time with conformally coupled fields, the ultraviolet subtracted one-point function \emph{vanishes} in the massless limit, thus both the self-generated mass term and the condensate are genuine radiative infrared effects in de Sitter space time and fields minimally coupled to gravity.

The self-consistent mass (\ref{selfmasa}) corresponds to
\be \Delta = \Big[ \frac{\lambda}{2\pi\,H}\Big]^\frac{2}{3}\,, \label{delfi3}\ee hence consistency with the approximation $\Delta \ll 1$ requires that $\lambda /H \ll 1$.

Furthermore, it follows from the above analysis that
\be \langle \phi^2 (\vec{x},\eta) \rangle = \frac{1}{C^2(\eta)}\Big[ \overline{\chi}^2(\eta)+ \mathcal{I}(\eta)\Big] \propto H^2 \, \Big[\frac{H}{\lambda}\Big]^{2/3}\,.\label{fi2cubi}\ee namely a de Sitter invariant result but non-analytic in the coupling.


\vspace{2mm}

\subsubsection{$\lambda \chi^4 $ theory:} 
\begin{figure}[h!]
\begin{center}
\includegraphics[height=1.5in,width=1.5in,keepaspectratio=true]{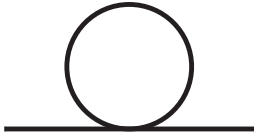}
\caption{Tadpole contribution to the self-energy in $\lambda\, \chi^4$. }
\label{fig:fi4selfenergytadpole}
\end{center}
\end{figure}

For $p=4$ the interaction Lagrangian density is $\mathcal{L}_I = -\lambda \,\chi^4$ the mechanism of self-consistent generation of mass via a tadpole contribution is implemented as follows: add a mass term to $\mathcal{L}_0$ and subtract it as a counterterm in $\mathcal{L}_I$ leading to
\be  \mathcal{L}_I     =     \frac{M^2}{2\,H^2\,\eta^2}   \; \chi^2 - {\lambda}  \;\chi^4 \,. \label{LIchi4} \ee Requiring that the counterterm cancels the contribution from the tadpole depicted in fig. (\ref{fig:fi4selfenergytadpole}) to the two-point function yields the condition
\be  \frac{M^2}{ H^2\,\eta^2} = 12 \,\lambda ~ \langle \chi^2(\vec{x},\eta) \rangle \label{tadcon}\ee which upon renormalization by subtracting the ultraviolet divergences and using the result (\ref{poletad}) leads to the self-consistent condition
\be \frac{M^2}{H^2\,\eta^2} = \frac{12\,\lambda }{8\pi^2\,\eta^2\,\Delta} \label{scfi4}\ee with the solution
\be M  = \Big[\frac{9}{2\pi^2}\Big]^\frac{1}{4}\, \lambda^\frac{1}{4}\,H \,.\label{scmassfi4}\ee The coupling dependence of this result is in agreement with those of refs.\cite{staroyoko,holman,serreau,garb}.

The self-consistent mass (\ref{scmassfi4}) leads to
\be \Delta = \Big[\frac{\lambda}{2\pi^2} \Big]^{\frac{1}{2}}\,, \label{delfi4}\ee and again, consistency with the approximation $\Delta \ll 1$ requires that $\lambda \ll 1$.

After the usual ultraviolet renormalization the equal time two point function of the unscaled scalar field (see eqn. (\ref{rescale}))  is de Sitter invariant, non-analytic in the coupling  and given by
\be \langle \phi^2(\eta) \rangle \propto \frac{H^2}{\sqrt{\lambda}} \label{fisq}\ee
which agrees with the result in refs.\cite{staroyoko,raja}.

\subsection{Normal ordering vs. self-consistent mass generation:} A simple approach would be to normal order the interaction Hamiltonian, which amounts to subtracting tadpole contributions thereby avoiding the problem of infrared and ultraviolet divergences. However, normal ordering requires specifying a vacuum state, as can be seen from the identity
\be :\chi_{\vec{k}}(\eta)\chi_{-\vec{k}}(\eta): = \chi_{\vec{k}}(\eta)\chi_{-\vec{k}}(\eta)-|g_\nu(k;\eta)|^2 \label{norord}\ee in particular normal ordering in the bare massless theory corresponds to $\nu =3/2$ leading to the infrared divergence of the tadpole, whereas normal ordering in the theory with a mass term in the non-interacting Lagrangian implies $\nu \simeq 3/2- \Delta$ and the infrared divergence is regularized, for conformally coupled massless particles   $\nu = 1/2$, and the tadpole only features the ultraviolet divergence of Minkowski space time but no infrared divergences.

 In the case of the cubic vertex for minimally coupled massless fields naive normal ordering neglects the fact that there is an infrared and secularly divergent radiatively induced vacuum expectation value, thus the vacuum state evolves to a  state in which the one point function is non-vanishing. Furthermore, this evolution of the expectation value is leading to the generation of a radiatively induced mass, which in turn regulates the infrared behavior. Similarly for the quartic vertex, the contribution from the tadpole and the self-consistent resummation entails that the vacuum state evolves on to another state in which the infrared divergences are regulated by the self-consistent mass.

Furthermore and important consequence is that the original massless theory breaks the underlying de Sitter invariance because of the infrared effects\cite{allen,folaci}, whereas the self-consistent generation of mass \emph{restores} de Sitter invariance by manifestly regularizing the infrared and leading to a time independent equal time
expectation value $\langle \phi^2(\eta) \rangle = \mathrm{constant}$, as befits de Sitter invariance. Thus the mechanism of self-consistent mass generation remarkably restores de Sitter invariance in the case of bare minimally coupled massless fields.

Last but not least, whereas normal ordering discards the tadpole divergences, the infrared divergences emerge in non-local self-energy contributions at one loop in the case of the cubic vertex or two loops in the case of the quartic vertex, this will be seen in detail below.

Through the self-consistent mass generation, the infrared divergences become poles in $\Delta$ and relieve the large logarithms at the expense of a   rearrangement of the perturbative expansion non-analytic in the coupling.

Thus normal ordering from the outset in the massless theory misses the important physics associated with the buildup of infrared effects, the fact that time evolution is rearranging the vacuum (and many particle states) in a manner that ultimately self-consistently regulates the infrared behavior. This situation is similar to the results from the stochastic approach\cite{staroyoko,richard}.

\vspace{2mm}

\subsection{Anomalous dimensions from self-consistent masses}
While the emergence of self-consistent masses has been previously recognized\cite{staroyoko,holman,leblond,rigo,garb}, it is noteworthy that this infrared mechanism of mass generation leads to \emph{anomalous scaling dimension} of the two point function in the superhorizon limit. In the bare massless theory the two point function scales as $1/k^3$ in this limit but after the self-consistent generation of mass these now behave as
\be \langle \chi_{\vec{k}}(\eta)\, \chi_{-\vec{k}}(\eta)\rangle \propto \frac{1}{k^{3-2\Delta}}\,. \label{anodimself}\ee where $\Delta$ is given by eqns. (\ref{delfi3},\ref{delfi4}) for cubic and quartic couplings respectively. We emphasize that the anomalous scaling dimension $\Delta$ is a non-perturbative result from the build-up of infrared fluctuations, much in the same way as anomalous scaling dimensions in the theory of critical phenomena. Furthermore, in the case of the cubic coupling, it is a result of the formation of the condensate, which is radiatively induced and a consequence of the infrared divergences of the massless theory.
The emergence of anomalous dimensions as a consequence of the infrared generated mass has also been recognized in ref.\cite{riotto}.

In the next section we show how the self-consistent method is systematically implemented within the Wigner-Weisskopf theory and also study how quantum states acquire a   decay width.

\section{Wigner-Weisskopf theory :}

\subsection{Transition amplitudes and probability}
In anticipation of the non-perturbative Wigner-Weisskopf theory in de Sitter cosmology and to identify corrections to masses and widths of the states, let us consider the example   of two interacting scalar fields, $\chi,\delta$ after conformal rescaling,  with interaction Hamiltonian
\be H_I = \frac{\lambda}{-H\,\eta} \int d^3 x ~ \chi(\vec{x},\eta)~ \delta^2(\vec{x},\eta)\,. \label{gencub}\ee The case $\delta = \chi$ is obtained straightforwardly. Using the expansion of the scalar field $\chi$ given by (\ref{quantumfieldds1}) and expanding the field $\delta$ as
\be \delta(\vec{x},\eta) = \frac{1}{\sqrt{V}}\sum_{\vec{k}} \Big[b_{\vec{k}}\,u_{\nubar}(k;\eta)\,e^{i\vec{k}\cdot\vec{x}}+ b^\dagger_{\vec{k}}\,\,u^*_{\nubar}(k;\eta)\,e^{-i\vec{k}\cdot\vec{x}}\Big]\,. \label{quantumfielddel}\ee where
\be u_{\nubar}(k,\eta) =  \frac{1}{2}\; i^{\nubar+\frac{1}{2}}
\sqrt{-\pi \eta}\,H^{(1)}_{\nubar}(-k\eta)\label{unubar}\ee with
\be {\nubar}^{\,2} =    \frac{9}{4}- \frac{M^2_{\delta}}{H^2}\,. \label{nubar}\ee The transition amplitude $\chi \rightarrow 2\delta$ is given by
\be \mathcal{A}_{\chi \rightarrow 2\delta}(\vec{k},\vec{p};\eta) =   \frac{2 i \,\lambda}{ H \,\sqrt{V}} \int^\eta_{\eta_0} \frac{d\eta_1}{\eta_1}\,g_{\nu}(k;\eta_1)\,u^*_{\nubar}(p;\eta_1)\,u^*_{\nubar}(q;\eta_1) ~~;~~q=|\vec{p}+\vec{k}| \label{amp}\ee and the total transition probability is
\be \mathcal{P}_{\chi \rightarrow 2\delta}(k;\eta) =    \int^\eta_{\eta_0}  {d\eta_2}  \int^\eta_{\eta_0} {d\eta_1} ~\Sigma(k\,;\eta_1,\eta_2) \label{proba}\ee where\footnote{A factor $1/2!$ accounts for Bose symmetry of the two particle final state.}
\be \Sigma(k\,;\eta_1,\eta_2) = \frac{2\,\lambda^2~g^*_\nu(k,\eta_2)\,g_\nu(k,\eta_1)}{H^2 \,\eta_1\,\eta_2} \,\int \frac{d^3p}{(2\pi)^3} ~ u^*_{\nubar}(p,\eta_1)\,u^*_{\nubar}(q,\eta_1)\,u_{\nubar}(p,\eta_2)\,u_{\nubar}(q,\eta_2)~~;~~q=|\vec{p}+\vec{k}| \label{selfe}\ee with the property that
\be \Sigma(k\,;\eta_2,\eta_1) = \Sigma^*(k\,;\eta_1,\eta_2)\,. \label{proper}\ee
Introducing the identity $1 = \Theta(\eta_2-\eta_1)+\Theta(\eta_1-\eta_2)$ in the (conformal) time integrals and using (\ref{proper}) we find
\be \mathcal{P}_{\chi \rightarrow 2\delta}(k;\eta) = 2   \int^\eta_{\eta_0}  {d\eta_2}  \int^{\eta_2}_{\eta_0}  {d\eta_1} ~\mathrm{Re}\Big[\Sigma(k\,;\eta_1,\eta_2) \Big] \label{proba2}\ee from which we identify the \emph{transition rate}
\be \Gamma(\eta) \equiv  \frac{d}{d\eta} \mathcal{P}_{\chi \rightarrow 2\delta}(k;\eta) =  2   \int^\eta_{\eta_0}  {d\eta'}   ~\mathrm{Re}\big[\Sigma(k\,;\eta ,\eta') \big] \label{gamma}\ee

\vspace{2mm}

The result can be extrapolated to the self-interacting $\lambda \chi^3$ case simply by replacing the mode functions $u_{\nubar}(k,\eta) \rightarrow g_\nu(k,\eta)$ and including the corresponding combinatoric factor.

In Minkowski space-time $\eta \rightarrow t$,   if the kinematics of the transition is allowed, namely energy-momentum conservation holds in the process, the transition is to on-shell states and the transition probability features a secular growth linear in time at long time, in which case the transition \emph{rate} becomes a constant. This is the result from  Fermi's Golden rule. If, on the other hand energy-momentum conservation is not fulfilled,   the  probability becomes constant at asymptotically long times, with a vanishing transition rate, describing virtual processes that contribute to wave function renormalization. A true decay of the quantum state is therefore reflected in a secular \emph{growth} of the transition probability and a transition rate that either remains constant or grows at asymptotically long time. In de Sitter space time the lack of a global time-like Killing vector implies the absence of kinematic thresholds and the lack of defined ``on-shell'' states. As discussed earlier in ref.\cite{boyan,boyprem} quanta of a single field can decay into other quanta of the same field, and more recently\cite{boyhol}  a generalization   of the non-perturbative Wigner-Weisskopf method to cosmology was introduced to study explicitly   the decay of quantum states.

\subsection{Wigner-Weisskopf theory in deSitter space time:}

In order to make the discussion self-contained, we highlight the main aspects of the Wigner-Weisskopf non-perturbative approach to study the decay of quantum states pertinent to the self-consistent description of mass generation discussed in the previous sections. For a more thorough discussion and comparison to results in Minkowski space time the reader is referred to ref.\cite{boyhol}.
 Expanding the interaction picture state $|\Psi(\eta)\rangle_I$ in   Fock states $|n\rangle$ obtained as usual by applying the creation operators on to the (bare) vacuum state (here taken to be the Bunch-Davies vacuum) as
  \be |\Psi(\eta)\rangle_I = \sum_n C_n(\eta) |n\rangle \label{expastate}\ee the evolution of the state in the interaction picture given by eqn. (\ref{ipds}) yields
  \be i \frac{d}{d\eta}|\Psi(\eta)\rangle_I = H_I(\eta)|\Psi(\eta)\rangle_I \label{eomip}\ee which in terms of the coefficients $C_n(\eta)$ become
   \be \frac{d\,C_n(\eta)}{d\eta}  = -i \sum_m C_m(\eta) \langle n|H_I(\eta)|m\rangle \,, \label{ecns}\ee it is convenient to separate the diagonal matrix elements, that represent \emph{local contributions}  from those that represent transitions and are associated with non-local self-energy corrections\footnote{In ref.\cite{boyhol} the diagonal matrix elements, hence the local contributions were not included.}, writing
    \be \frac{d\,C_n(\eta)}{d\eta}  = -i C_n(\eta)\langle n|H_I(\eta)|n\rangle -i \sum_{m\neq n} C_m(\eta) \langle n|H_I(\eta)|m\rangle \,. \label{ecnsoff}\ee
    Although this equation is exact, it yields   an infinite hierarchy of simultaneous equations when the Hilbert space of states $|n\rangle$ is infinite dimensional. However, progress is made by considering the transition between states connected by the interaction Hamiltonian at a given order in $H_I$:
consider the case when one state, say $|A\rangle$ couples to a set of states $|\kappa\rangle$, which couple back to $|A\rangle$ via $H_I$, to lowest order in the interaction the system of equation closes in the form
 \bea \frac{d\,C_A(\eta)}{d\eta} & = & -i   \langle A|H_I(\eta)|A\rangle \, C_A(\eta)-i \sum_{\kappa \neq A} \langle A|H_I(\eta)|\kappa\rangle \,C_\kappa(\eta)\label{CA}\\
\frac{d\,C_\kappa(\eta)}{d\eta}& = & -i \, C_A(\eta) \langle \kappa|H_I(\eta) |A\rangle \label{Ckapas}\eea where the $\sum_{\kappa \neq A}$ is over all the intermediate states coupled to $|A\rangle$ via $H_I$ representing transitions.

Consider the initial value problem in which at time $\eta=\eta_0$ the state of the system is given by $|\Psi(\eta=\eta_0)\rangle = |A\rangle$ so that \be C_A(\eta_0)= 1 ~~;~~ C_{\kappa\neq A}(\eta=\eta_0) =0 \,,\label{initial}\ee  solving (\ref{Ckapas}) and introducing the solution into (\ref{CA}) we find \bea  C_{\kappa}(\eta) & = &  -i \,\int_{\eta_0}^{\eta} \langle \kappa |H_I(\eta')|A\rangle \,C_A(\eta')\,d\eta' \label{Ckapasol}\\ \frac{d\,C_A(\eta)}{d\eta} & = & -i   \langle A|H_I(\eta)|A\rangle \, C_A(\eta) - \int^{\eta}_{\eta_0} \Sigma_A(\eta,\eta') \, C_A(\eta')\,d\eta' \label{intdiff} \eea where\footnote{In ref.\cite{boyhol} it is proven that in Minkowski space-time the self-energy in the single particle propagator is given by $i\Sigma$.} \be \Sigma_A(\eta,\eta') = \sum_{\kappa \neq A} \langle A|H_I(\eta)|\kappa\rangle \langle \kappa|H_I(\eta')|A\rangle \,. \label{sigma} \ee


Equation (3.19) makes manifest that $\Sigma_A(\eta,\eta')$ is the \emph{retarded} self-energy,  since $\eta > \eta'$. In ref.\cite{boyhol} the correspondence between the Wigner-Weisskopf and the Dyson resummation of the propagator in Minkowski space-time is explained in detail and shown that the self-energy that enters in the evolution equation for the amplitudes is the retarded one, resulting in that the propagator features a pole in the upper half of the complex frequency plane (see eqns. (2.12), (2.30)  in ref.\cite{boyhol}). The retarded self-energy is the one that enters in the description of an initial value problem as befits the evolution of the amplitudes from an  initial state.


In eqn. (\ref{Ckapas}) we have not included the diagonal term as in (\ref{CA})\footnote{These diagonal terms represent local self-energy insertions in the propagators of the intermediate states, hence higher orders in the perturbative expansion.}, it is clear from (\ref{Ckapasol}) that with the initial condition (\ref{initial}) the amplitude of $C_{\kappa}$ is of $\mathcal{O}(H_I)$ therefore a diagonal term would effectively lead to higher order contributions to (\ref{intdiff}). The integro-differential equation  (\ref{intdiff}) with \emph{memory} yields a non-perturbative solution for the time evolution of the amplitudes and probabilities, which simplifies in the case of weak couplings. In perturbation theory   the time evolution of $C_A(\eta)$ determined by eqn. (\ref{intdiff}) is \emph{slow} in the sense that
the time scale is determined by a weak coupling kernel $\Sigma_A$, hence an approximation in terms of an expansion in derivatives of $C_A$ emerges as follows: introduce
\be W (\eta,\eta') = \int^{\eta'}_{\eta_0} \Sigma_A(\eta,\eta'')d\eta'' \label{Wo}\ee so that \be \Sigma_A(\eta,\eta') = \frac{d}{d\eta'}\,W (\eta,\eta'),\quad W (\eta,\eta_0)=0. \label{rela} \ee Integrating by parts in eq.(\ref{intdiff}) we obtain \be \int_{\eta_0}^{\eta} \Sigma_A(\eta,\eta')\,C_A(\eta')\, d\eta' = W (\eta,\eta)\,C_A(\eta) - \int_{\eta_0}^{\eta} W (\eta,\eta')\, \frac{d}{d\eta'}C_A(\eta') \,d\eta'. \label{marko1}\ee The second term on the right hand side is formally of \emph{higher order} in $H_I$, integrating by parts successively   yields  a systematic approximation scheme as discussed in ref.\cite{boyhol}.

 Therefore to leading order in the interaction we find
 \be C_A(\eta) = e^{-\int^{\eta}_{\eta_0}\widetilde{W} (\eta',\eta')\, d\eta'}  , \quad \widetilde{W} (\eta',\eta')= i   \langle A|H_I(\eta')|A\rangle + \int_{\eta_0}^{\eta'} \Sigma_A(\eta',\eta^{''}) d\eta^{''}\,. \label{dssolu} \ee

  This expression has a clear and simple intepretation in Minkowski space time by replacing $\eta,\eta' \rightarrow t,t'$ with  the states $|n\rangle$ being eigenstates of the unperturbed Hamiltonian $H_0$,  and $H_I(t) = e^{i H_0 t}H_I(0)e^{-i H_0 t}$: the contribution $\langle A|H_I(t')|A\rangle=\langle A|H_I(0)|A\rangle =\delta E^{(1)}_A $ is simply the first order shift in the energy in elementary quantum mechanics and   taking the long time limit $t\rightarrow \infty$ (with a convergence factor $i\epsilon$)
  \be \int^{t\rightarrow \infty}_0\Sigma_A(t,t')dt' = i\sum_{\kappa \neq A} \frac{|\langle A|H_I|\kappa\rangle |^2}{ E_\kappa -E_A +i\epsilon} \equiv~ i\,\delta E^{(2)}_A + \frac{\Gamma}{2} \ee   thus the imaginary part of the time integral yields the second order energy shift and the real part yields half of the decay rate \emph{ a la} Fermi's golden rule. Inserting this result in (\ref{dssolu}) yields, in Minkowski space-time
  \be C_A(t) = e^{-i \delta E_A\,t}~ e^{-\Gamma t/2} ~~;~~ \delta E_A = \delta E^{(1)}_A + \delta E^{(2)}_A \,. \label{CAminko}\ee
  A more careful treatment of the long time limit also exhibits the wave function renormalization\cite{boyhol} and   establishes the equivalence with the dynamical renormalization group\cite{drg} in Minkowski space time.

  Motivated by this interpretation we introduce the \emph{real quantities} $\mathcal{E}_A(\eta)\,;\,\Gamma_A(\eta)$ as
 \be  i   \langle A|H_I(\eta')|A\rangle +\int^{\eta'}_{\eta_0} \Sigma_A(\eta',\eta'')d\eta'' = i\,\mathcal{E}_A(\eta')+ \frac{1}{2}~\Gamma_A(\eta') \label{realima}\ee in terms of which
 \be  C_A(\eta) = e^{-i\int^{\eta}_{\eta_0}\mathcal{E}_A(\eta') d\eta'}~ e^{-\frac{1}{2}\int^{\eta}_{\eta_0}\Gamma_A(\eta') d\eta'} \label{caofeta}\ee When the state $A$ is a single particle state, radiative corrections to the mass are extracted from $\mathcal{E}_A$ and
 \be \Gamma_A(\eta) = -  \frac{d}{d\eta}\ln\Big[|C_A(\eta)|^2\Big] \label{decarate} \ee is identified as a (conformal) time dependent decay rate. We see from (\ref{decarate}) that $\Gamma(\eta)$ is exactly the same as expression (\ref{gamma}).

 In Minkowski space-time $\mathcal{E}_A$ corresponds to the self-energy correction to the mass of the particle\cite{boyhol,qoptics} and the program of renormalized perturbation theory begins by writing the free field part of the Lagrangian in terms of the renormalized mass and introducing a counterterm in the interaction Lagrangian so that it cancels the radiative corrections to the mass from the self-energy. Namely the counterterm in the interaction Lagrangian is fixed by requiring that for the single particle state $|A\rangle = |1_{\vec{k}}\rangle$, in the long time limit $\eta' \rightarrow 0^-$
 \be \mathcal{E}_{1\vec{k}}(\eta') =\langle 1_{\vec{k}}|H_I(\eta')|1_{\vec{k}}\rangle + \int_{\eta_0}^{\eta'} \mathrm{Im}\Big[\Sigma_1(k;\eta',\eta^{''})\Big] d\eta^{''} = 0\,. \label{renmass}\ee  We will implement the \emph{same strategy} to obtain the self-consistent radiatively generated mass in de Sitter space time where equation (\ref{renmass}) will determine the \emph{self-consistent condition} for the mass.
 In Minkowski space time, the condition (\ref{renmass}) is tantamount to requiring that the (real part of the) pole in the propagator be at the physical mass\cite{boyhol}.

 \vspace{1cm}

 \subsection{Cubic vertex:}

 We now   consider the cubic interaction given by the Lagrangian density (\ref{LI}).   In the interaction picture the fields are expanded as in (\ref{quantumfieldds1}),  to carry out the perturbative expansion in the state in which the expectation value of the field vanishes, the  field $\chi$ is shifted by its vacuum expectation value as in (\ref{shift}), and include the tadpole counterterm given by the last term in  eqn. (\ref{shiftedLI}), thus replacing the interaction Hamiltonian   by that obtained from (\ref{shiftedLI}) and treating the quadratic term $\Psi^2$ as a mass counterterm, namely from eqn. (\ref{shiftedLI})
 \be  {H}_I(\eta) = \int d^3 x  \Bigg\{ \frac{\delta M^2}{2\,H^2\,\eta^2 }   \Psi^2 ~ -\frac{\lambda}{H\,\eta } :\Psi^3: - \Bigg(\mathcal{J}(\eta) +  \frac{3\lambda}{ H\,\eta} \; \Big[\mathcal{I}(\eta)+ \overline{\chi}^{\,2}(\eta)\Big]\Bigg)\Psi \Bigg\}\,.\label{shiftedHIpsi}\ee  The   counterterm $\mathcal{J}(\eta)$ is required to cancel the linear term in $\Psi$ in (\ref{shiftedHIpsi}), leading to the relation (\ref{onelupcount}),  simultaneously ensuring that   \be \langle 0| \Psi |0\rangle = 0~~;~~\langle 1_{\vk=\vec{0}}|H_I(\eta)|0 \rangle = 0 \,. \label{novac}\ee

  We focus on the time evolution of single particle states of the field $\chi$, namely $|1_{\vk}\rangle$. The mass counterterm contributes only to the diagonal matrix element
 \be \langle 1_{\vk}|H_I(\eta)|1_{\vk} \rangle = \frac{\delta\,M^2 }{ H^2\,\eta^2}\,|g_\nu(k,\eta)|^2\, ~~;~~ \frac{\delta\,M^2 }{H^2} =  \frac{3\,\lambda^2}{4\pi^2 \,H^2 \Delta^2}-\frac{M^2 }{H^2}\,,\label{diagME}\ee

\begin{figure}[h!]
\begin{center}
\includegraphics[height=3in,width=4in,keepaspectratio=true]{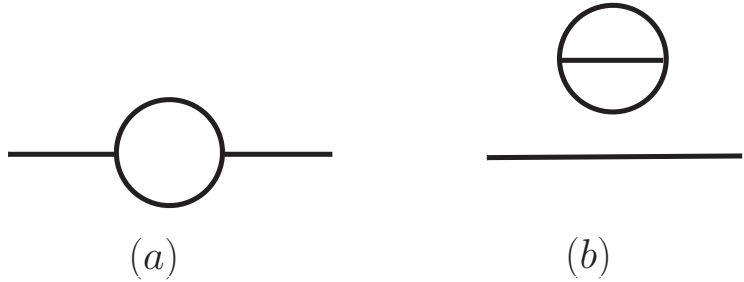}
\caption{Lowest order contributions to the evolution of the single particle states.   Fig. (a): irreducible self energy, fig. (b): vacuum diagram.}
\label{fig:se3}
\end{center}
\end{figure}

 The interaction Hamiltonian connects the state $|1_{\vk}\rangle$ to an intermediate state with two particles $|1_{\vec{p}};1_{\vec{k}-\vec{p}}\rangle$ and also to the  state $|1_{\vk};1_{\vq};1_{\vec{p}};1_{\vec{k}-\vec{q}}\rangle$, the first state
 is connected back to $|1_{\vk}\rangle$ by $H_I$ in the self-energy contribution depicted in fig.(\ref{fig:se3}-a)  whereas the second state contributes to the vacuum disconnected diagram displayed in the same figure (\ref{fig:se3}-b). The vacuum diagram is subtracted out consistently in perturbation theory by redefining the dressed states constructed out of the dressed vacuum. See discussion in ref.\cite{boyhol}.

The irreducible self-energy diagram (a) is given by
\be  \Sigma(k,\eta,\eta')=\frac{18\lambda^2}{ H^2\,\eta\,\eta'}~ g^*_\nu(k;\eta)g_\nu(k;\eta')  \int \frac{d^3p}{(2\pi)^3}~ g_{\nu }(p;\eta)\,g^*_{\nu }(p;\eta')\,g_{\nu }(|\vec{k}-\vec{p}|;\eta)\,g^*_{\nu }(|\vec{k}-\vec{p}|;\eta')
\label{sefull}\ee with  $g_\nu(k;\eta)$ given by eqns. (\ref{gnu}). We recognize eqn. (\ref{sefull}) as the kernel in the transition probability (\ref{proba},\ref{selfe}) with the only difference being the combinatoric factors and the replacement $ u_{\overline{\nu}} \rightarrow g_\nu $.
Furthermore with $G(k,\eta,\eta') = g^*_\nu(k;\eta)g_\nu(k;\eta')$ it follows that
\be \int^\eta_{\eta_0} \Sigma(k,\eta,\eta')~d\eta' \propto \int \frac{d^3p}{(2\pi)^3} ~\int \frac{d\eta'}{H\eta'} ~ G(k,\eta,\eta')\,G(p,\eta,\eta')\,G(|\vec{k}-\vec{p}|,\eta,\eta') \label{bispec}\ee and the imaginary part of the $\eta'$-integral is proportional to  the \emph{bispectrum} of the scalar field\cite{nongauss}, an observation that will become important in the interpretation of the final result below.

To simplify notation we introduce
\be H^{(1)}_{\nu }(-p \,\eta)H^{(2)}_{\nu }(-p \, \eta') \equiv \mathcal{H}_{\nu }(p;\eta;\eta') \label{capHdef}\ee and carry out the angular integral by changing variables to
\be q= \Big[k^2+p^2+2kp\cos(\theta)\Big]^\frac{1}{2} \Rightarrow d\cos(\theta) = \frac{q}{k\,p}\,dq \label{change}\ee
in terms of which the self-energy is given by
\be  \Sigma(k;\eta;\eta')    =    \frac{9\, \lambda^2}{32\,H^2\,k}~ g^*_\nu(k;\eta)g_\nu(k;\eta') ~I [k,\eta,\eta']\,, \label{selfa} \ee  where
\be I [k,\eta,\eta'] =
 \int_0^\infty \mathcal{H}_{\nu }(p;\eta;\eta')~\Big[\mathcal{F}_{\nu }(k+p;\eta\;\eta')-\mathcal{F}_{\nu }(|k-p|;\eta\;\eta')\Big] p\,dp
\label{Inu}\ee and\cite{grad}
\be \mathcal{F}_{\nu }(q;\eta\;\eta') = \int  \mathcal{H}_{\nu }(q;\eta;\eta')~ q\,dq = \frac{(-q\eta')\,H^{(1)}_{\nu }(-q\eta)\,H^{(2)}_{\nu-1}(-q\eta')-(-q\eta)\,H^{(2)}_{\nu }(-q\eta')\,H^{(1)}_{\nu -1}(-q\eta) }{\eta^2-{\eta'}^{2}}\,. \label{calFdef}\ee The integrals feature infrared divergences at $p \rightarrow 0$ and $|k-p| \rightarrow 0$ a result that follows from the identities
\bea &&  \mathcal{H}_{\nu }(p;\eta;\eta')~~  \stackrel{p\rightarrow 0}{=}~~ \Big[\frac{\Gamma(\nu)}{\pi} \Big]^2\, \Big(\frac{4}{\eta\,\eta'}\Big)^{\nu}~p^{-2\nu } \label{capHsmallp} \\ &&
\mathcal{F}_{\nu }(q;\eta;\eta') ~~  \stackrel{q\rightarrow 0}{=}~~  -\frac{\Gamma(\nu )\,\Gamma(\nu-1)}{2\pi^2}\, \Big(\frac{4}{\eta\,\eta'}\Big)^{\nu}~q^{2-2\nu} \label{calFsmallq}\eea
The integrals near the region $p\simeq 0$ and $p \simeq k$ yield simple poles in
\be  {\Delta} = \frac{3}{2}-\nu \label{deltabar}\,.\ee In the appendix we carry out the integrals keeping the poles in $ {\Delta}$ and the leading infrared logarithmic contribution. The final result for $I[k,\eta,\eta']$ in eqn. (\ref{selfa}) up to the leading and next to leading order in $ {\Delta}$ is  given by eqn. (\ref{Ifinali}).

The first term in the first line of eqn. (\ref{Ifinali}) is the leading order contribution in the $\Delta \rightarrow 0$ limit and encodes the leading infrared behavior. The second term  in the first line is recognized as the contribution from a massless particle conformally coupled to gravity and yields the leading contribution for modes deep inside the Hubble radius, since for these modes the behavior of the Hankel functions is described by the simple Minkowski space time plane waves\cite{boyhol}.  These two terms yield
\bea  \Sigma(k,\eta,\eta') & = & \frac{9 \lambda^2}{2H^2\pi^2}\,\frac{|g_\nu(k,\eta)|^2\,|g_\nu(k,\eta')|^2}{(\eta \eta')^2~\Delta}\big[k^2\eta\eta' \big]^{\Delta} \nonumber\\ & + &
\frac{9 \lambda^2}{8H^2\pi^2}\,g^*_\nu(k;\eta)g_\nu(k;\eta')\,\Bigg[ \frac{e^{-ik(\eta-\eta')}}{i(\eta \eta')} ~ \mathcal{P}\Bigg(\frac{1}{\eta-\eta'}\Bigg) + \frac{\pi}{\eta^2}~\delta(\eta-\eta') \Bigg] \,.\label{twoterms}\eea Although the term in the second line is subleading in $\Delta$, we have included it because it yields the leading contribution from \emph{subhorizon modes} which are completely described by a conformally coupled massless field and contain the short distance ultraviolet divergence associated with the usual mass renormalization.

From the result eqn. (\ref{secline}), it follows that
in the integral
\be W(\eta,\eta) = \int^{\eta}_{\eta_0}   \Sigma(k,\eta,\eta')~d\eta' \ee
the second term  in (\ref{twoterms}) yields the contribution found in ref.\cite{boyhol} in the case of massless conformally coupled particles
\be W_{cc}(\eta,\eta)= \frac{9\,\lambda^2\,\big| g_\nu(k,\eta)\big|^2}{8 \pi^2 H^2\eta^2} \Big[\frac{\pi}{2}+ ~i\,\ln\big(\tilde{\epsilon}\big) \Big] \label{confcont}\ee The imaginary part is an ultraviolet divergent contribution to the mass, which is canceled by the mass counterterm (\ref{diagME}) whereas the real part gives the contribution to the decay width found in ref.\cite{boyhol} for conformally coupled massless particles, which, however is subleading in $\Delta$. The first term is purely real and does \emph{not} contribute to the mass thus the results (\ref{diagME}) and (\ref{confcont}) combined with the condition (\ref{renmass}) leads to
\be  \frac{|g_\nu(k,\eta)|^2 }{ \eta^2}~\Big[ \frac{\delta M^2 }{H^2}+ \frac{9\,\lambda^2\,}{8 \pi^2 H^2} \ln\big(\tilde{\epsilon}\big) \Big]=0\,.\label{mascondi}\ee
 Absorbing the ultraviolet divergence in a renormalization of the mass $M^2_{ ren}$ the self-consistent condition becomes \be \frac{M^2_{ ren}}{H^2}= \frac{3\,\lambda^2}{4\pi^2 \,H^2 \Delta^2_{ren}} \label{renoscmas}\ee which yields
 \be \Delta_{ren} = \Big[\frac{\lambda}{2\pi H} \Big]^\frac{2}{3}~~;~~  M_{ren} = H~\sqrt{3}\,\Big[\frac{\lambda}{2\pi H}\Big]^{1/3}  \,. \label{selfmasaren}\ee confirming the result (\ref{selfmasa}) in the previous section but now including the short distance renormalization. In what follows we will refer to $\Delta_{ren}, M_{ren}$ simply as $\Delta, M $ with the implicity understanding of renormalized quantities.

Although the self-consistent mass (\ref{selfmasaren}) is similar to the result in ref.\cite{rigo}, we note that our result reveals that the irreducible self-energy features a \emph{single pole} in $\Delta$, in agreement with the perturbative study in ref.\cite{smit} and the ``infrared counting'' of ref.\cite{holman} but in disagreement with the results of the plane wave \emph{ansatz} proposed in ref.\cite{rigo}\footnote{The author has not been able to understand the source of the discrepancy, the plane wave \emph{ansatz} proposed in ref.\cite{rigo} does not seem to include the non-locality in conformal time explicit in the result (\ref{twoterms}) and that of ref.\cite{smit}.}
   and that the self-consistent mass arises from the tadpole (expectation value), it is the result of the formation of the infrared condensate,  and \emph{not} from the irreducible self-energy which is  \emph{real} to leading order in $\Delta$. These differences notwithstanding, the self-consistently generated mass features the coupling and $H$ dependence in agreement with the results of ref.\cite{rigo} with a slightly different factor.

   \vspace{2mm}

\textbf{Superhorizon modes:} For $-k\eta, -k\eta' \ll 1$ the first line in eqn. (\ref{twoterms}) dominates and
\be  \Sigma(k,\eta,\eta') \simeq \frac{9\lambda^2\,k^2}{32H^2\Delta}\,\frac{|H^{(1)}_\nu(-k\eta)|^2}{ [-k\eta ]^{1-\Delta}} ~~ \frac{|H^{(1)}_\nu(-k\eta')|^2}{ [-k\eta' ]^{1-\Delta}}\,. \label{suphor}\ee Therefore requiring the condition (\ref{mascondi}) we find from the results (\ref{dssolu},\ref{caofeta})
\be  \widetilde{W}(\eta,\eta)= \frac{\Gamma(\eta) }{2} = \frac{9\lambda^2\,k}{32H^2\Delta}~\frac{|H^{(1)}_\nu(z)|^2}{z^{1-\Delta}} ~~\int_z^{z_0} \frac{|H^{(1)}_\nu(z')|^2}{(z')^{1-\Delta}} ~~dz' \label{intesigi}\ee where $z=-k\eta,z_0 = -k\eta_0$. The integral in (\ref{intesigi}) is remarkably similar to an expression that emerges in the two point correlator in refs.\cite{kroto,leblond} upon the analytic continuation $\nu  \rightarrow -i\mu$ where $\mu = [\frac{M^2}{H^2}-\frac{9}{4}]^\frac{1}{2} >0$ and real in those references. However for $\nu = -i\mu$ as in \cite{kroto,leblond} the limits of the integral can be taken $z_0 \rightarrow \infty, z\rightarrow 0$, whereas the lower limit cannot be taken to vanish for $\nu \sim 3/2 -\Delta$ and real because of the strong infrared divergence, a consequence of light masses. This is an important difference with the case studied in these references that prevents a meaningful comparison.

  We can now obtain $\widetilde{W}(\eta,\eta)$ in eqn. (\ref{dssolu}), since  $\Sigma$ is real to leading order in $\Delta$, requiring that the tadpole cancels the imaginary part of $\widetilde{W}$ and carrying out the remaining integrals in the limit $z \ll 1~;~z_0 \simeq 1$ (when the particular mode crosses the Hubble radius) we finally  find
\be  C_{1k}(\eta)  \simeq e^{-\gamma(-k\eta)} ~~;~~ \gamma(-k\eta)= \frac{\lambda^2}{16\pi^2H^2 \Delta} [-k\eta]^{-6(1-\Delta)}\,. \label{C2}\ee Using the result (\ref{selfmasaren})  we find
\be \gamma(-k\eta) = \frac{1}{4}\,\Big(\frac{\lambda}{2\pi H}\Big)^\frac{4}{3} \Big[\frac {H}{k_{ph}(\eta)}\Big]^{6(1-\Delta)}+\cdots\,. \label{Csupfina}\ee The dots stand for higher order terms suppressed with respect to the leading term by at least a factor $\big[\lambda/H \big]^{2/3}$. Thus we see that the self-consistent mass generation leads to a \emph{rearrangement of the perturbative expansion non-analytic in the coupling $\lambda$}.

\begin{figure}
\begin{center}
\includegraphics[height=4in,width=4in,keepaspectratio=true]{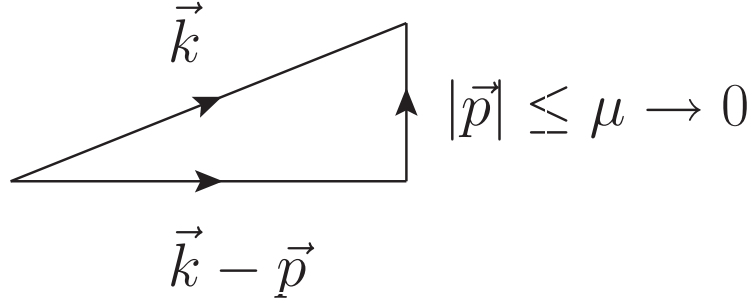}
\caption{Triangle of momenta for the bispectrum (see eqn.(\ref{bispec})) integration over $|\vec{p}|< \mu \rightarrow 0$ corresponds to the highly squeezed limit and yields the pole in $\Delta$.}
\label{triangle}
\end{center}
\end{figure}

The strong suppression of the single particle amplitude for subhorizon modes is a consequence of emission and absorption of soft superhorizon quanta. The dependence on the wavevector $\propto 1/k^6$ has a simple interpretation in terms of the relation of $\Sigma(k,\eta,\eta')$ with the bi-spectrum\cite{nongauss} highlighted by eqn. (\ref{bispec}): the pole in $\Delta$ arises from the integration in a small band   $|\vec{p}| < \mu \rightarrow 0$ of the highly squeezed triangle for the bispectrum configuration displayed in fig. (\ref{triangle}). The extra power $6\Delta$ in (\ref{Csupfina}) reflects the anomalous dimension.

From the result for the self-energy obtained in the appendix it is clear to next order in $\Delta \propto \big[\lambda/H\big]^{2/3}$ that there are logarithmic corrections $\propto \Delta\, \ln[-k\eta]$  to $\gamma(-k\eta)$   thus these corrections will become relevant when $-k\eta \simeq e^{1/\Delta}$ when the single particle amplitude is already strongly suppressed. These  corrections may be included  in the self-energy and the non-perturbative Wigner-Weisskopf resummation automatically exponentiates these potentially secular terms  into a correction to  the exponent $\gamma(-k\eta)$ in (\ref{C2}). Clearly extra logarithms in higher orders in $\Delta$ only modify the decay law but their ``secularity'' is automatically resummed into the overall decay function $\gamma(-k\eta)$.

\vspace{2mm}

\textbf{Subhorizon modes:}  For $-k\eta_0 \gg -k\eta, -k\eta' \gg 1$ close inspection of the self-energy contributions given by eqn. (\ref{Ifinali}) shows that only the first two terms displayed in the first line of the equation are dominant, however the contribution from $\mathcal{D}[k,\eta,\eta']$ is suppressed with respect the term $1/\Delta + \ln[-k\eta]$ in the bracket. The second term in the first line is given by eqn. (\ref{secline}).
Furthermore, in the subhorizon regime
\be g^*_\nu(k;\eta)g_\nu(k;\eta') = \frac{e^{ik(\eta-\eta')}}{2k} \label{subho}\ee after absorbing the short distance divergence in a renormalization of  the mass and imposing the condition (\ref{renoscmas}) we find
\be W(\eta,\eta) = \frac{9\lambda^2\,k}{8  \pi^2 H^2 \Delta} \Bigg[ \frac{1}{z^{3-\Delta}}+ \frac{\pi \,\Delta}{4\,z^2 }\Bigg]~~;~~z=-k\eta =k_{ph}(\eta)/H \,.   \label{subho}\ee  The final integral is performed in the limit $-k\eta_0 \gg -k\eta \gg 1$ and using (\ref{selfmasaren}) leads to the final result for sub-horizon modes

 \be  C_{1k}(\eta)  =   e^{-\gamma(\eta)} ~~;~~ \gamma(\eta) = \frac{9}{4}\, \Big( \frac{\lambda}{2\pi\,H}\Big)^\frac{4}{3}\,\Big[\frac{H}{k_{ph}(\eta)}\Big]^{2-\Delta}\,\Bigg[1+\frac{\pi}{2} \, \Big( \frac{\lambda}{2\pi\,H}\Big)^\frac{2}{3}\,\frac{k_{ph}(\eta)}{H}  \Bigg] +\cdots   \label{subdecay} \ee

 This expression highlights both the non-perturbative nature of the expansion in terms of fractional powers of $\lambda/H$ as a result of the infrared divergences and the limit of validity of the leading order approximation determined by the pole in $\Delta$ for subhorizon modes: the second term in the bracket begins to dominate for
 \be \frac{k_{ph}(\eta)}{H} > \Big(\frac{2\pi\,H}{\lambda}\Big)^\frac{2}{3}\,.\label{breakapx}\ee

 \vspace{2mm}

\begin{figure}[ht!]
\begin{center}
\includegraphics[height=4in,width=4in,keepaspectratio=true]{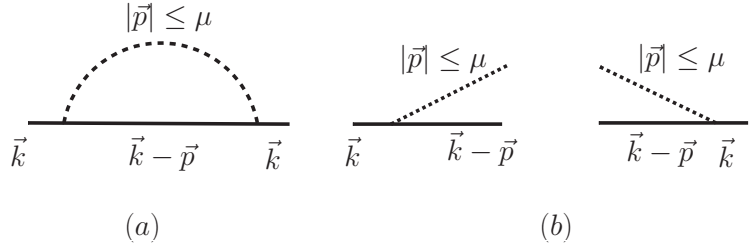}
\caption{Processes that contribute to the leading order poles in $\Delta$: (a) intermediate state of superhorizon modes, (b) emission and absorption of superhorizon quanta.}
\label{figsoftse}
\end{center}
\end{figure}

 \textbf{Corollary and rules to extract the poles in $\Delta $:} From the explicit calculation in the appendix and the results obtained above we learn that the leading infrared behavior manifest as the leading poles in $\Delta$ correspond to the process of absorption and emission of superhorizon quanta depicted in fig. (\ref{figsoftse}) for the self-energy for the cubic interaction. The pole arises from the integration in a small band of superhorizon wavevectors of width $\mu \rightarrow 0$ and can be extracted by implementing a set of simple rules that can be inferred from the example of the one-loop diagram corresponding to the self-energy in the theory with cubic interaction depicted in fig. (\ref{fig:se3})  and given by eqn. (\ref{sefull}). The momentum integral features two regions in which the infrared behavior leads to poles in $\Delta$, $p < \mu$ and $|k-p|<\mu$,  however, the second region is equivalent to the first by  rerouting the external momentum, this yields an overall factor $2$, namely the number of possibilities to reroute the external momenta along one internal lines.  In the region of momentum integration $p\leq \mu$ we can set $|k-p| \sim k$ and extract the product $ g_{\nu}(|\vec{k}-\vec{p}|,\eta)\,g^*_{\nu}(|\vec{k}-\vec{p}|,\eta') \sim g_{\nu}(k,\eta)\,g^*_{\nu}(k,\eta')$ outside the integral which becomes
 \bea \frac{1}{2\pi^2} \int^\mu_0 p^2~  g_{\nu}(p;\eta)\,g^*_{\nu}(p;\eta')~ dp & = &  \frac{1}{8\pi}~\sqrt{\eta\,\eta'}~~   \frac{\Gamma^2(\nu)}{\pi^2}  ~ \Big(\frac{4}{\eta\,\eta'}\Big)^{\nu}~ \frac{\mu^{2\Delta}}{2  {\Delta}} \nonumber \\ & \simeq & \frac{1}{8\pi^2 \Delta\,\eta\,\eta'}\,\Big[1+\frac{\Delta}{2}\,\ln[\mu^2\eta\eta']+\cdots\Big]
   \label{softin}\eea and the $\ln[\mu]$ is cancelled by the integration with $p > \mu$ as explicitly shown in the appendix.

  Thus to leading order in  $ {\Delta}$ the self-energy (\ref{sefull}) becomes
 \bea \Sigma_{lo}(k,\eta,\eta') & = &  \frac{18\lambda^2}{H^2\,\eta \, \eta'}~ g^*_\nu(k;\eta)g_\nu(k;\eta') \,(2)\, \Bigg[g_{\nu}(k,\eta)\,g^*_{\nu}(k,\eta') \Bigg]~\Bigg[ \frac{1}{8\pi^2 \Delta\,\eta\,\eta'}\Bigg]+\cdots \nonumber \\ & = & \frac{9\lambda^2}{2 \pi^2 \,H^2\, \Delta}~
 \frac{|g_\nu(k;\eta)|^2 \, |g_\nu(k;\eta')|^2}{\big(  \eta\,\eta' \big)^2  }   +\cdots \label{leador}\eea the first term $g^*_\nu(k;\eta)g_\nu(k;\eta')$ corresponds to the external lines, the factor $(2)$ is the number of lines through which the external momentum $k$ can be rerouted, the first bracket corresponds to the internal line that carries the external momentum $g_{\nu}(k;\eta)\,g^*_{\nu}(k;\eta') $, the last bracket is the leading order contribution from the integration of superhorizon modes in the band $0\leq p\leq \mu \rightarrow 0$ and the dots stand for subleading terms that do not feature   poles in ${\Delta}$. The expression (\ref{leador}) confirms the leading order result of the self-energy (\ref{twoterms}).

 This result generalizes to the following set of rules valid for an irreducible self-energy graph with $n+1$ internal lines and $n$ loop integrations
 \begin{itemize}
 \item{ Reroute the external momenta to run along one of the internal lines in the graph: there are $n+1$ possibilities that give the overall factor $n+1$, the ``propagator'' associated with this internal line is
     \be  g_{\nu}(k,\eta)\,g^*_{\nu}(k,\eta')\,,\label{intprop}\ee}
 \item{For each loop momentum integral (with $0 \leq p \leq \mu$) associate a factor \be \frac{1}{8\pi^2\, \Delta\,(\eta\,\eta')} \,,\label{looprule}\ee arising from the integration within a band of width $\mu \rightarrow 0$ of superhorizon quanta. Therefore the order of the pole in $\Delta$ is determined by the number of soft internal lines, hence the number of  superhorizon quanta emitted by the decaying particle as depicted in fig. (\ref{figsoftse}). }

     \item{The external lines correspond to \be g^*_\nu(k;\eta)g_\nu(k;\eta')\,.\label{extlin}\ee }
     \end{itemize}

\subsection{Quartic self-interaction:}\label{subsec:quartic} We are now in position to use the above rules to obtain the leading order contribution to the self-consistent mass and decay width for the case of $\lambda \chi^4$, whose interaction Hamiltonian is given by
\be  H_I     =  \int d^3 x ~  \Bigg[-\frac{M^2}{2\,H^2\,\eta^2}\, \chi^2 + {\lambda}  \;\chi^4 \Bigg]\,. \label{HIchi4} \ee

 The one loop tadpole diagram   shown in fig. (\ref{fig:fi4selfenergytadpole}) contributes to the diagonal matrix element
\be \langle 1^\chi_{\vk}|H_I(\eta)|1^\chi_{\vk} \rangle = \frac{\delta\,M^2 }{ H^2\,\eta^2}\,|g_\nu(k,\eta)|^2\, ~~;~~ \frac{\delta\,M^2 }{H^2} =  \frac{12\,\lambda }{8\pi^2\, \Delta}  -\frac{M^2 }{H^2}\,.\label{diagMEfi4}\ee

\begin{figure}[ht!]
\begin{center}
\includegraphics[height=2in,width=2in,keepaspectratio=true]{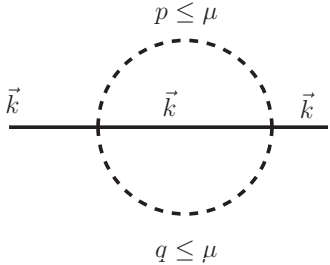}
\caption{Leading order in $\Delta$ for the two loop self-energy in $\lambda \chi^4$. The dashed lines correspond to the integration of the superhorizon modes within a band of width $\mu \rightarrow 0$.   }
\label{figsoftsefi4}
\end{center}
\end{figure}

We obtain the leading order infrared pole of the two loop contribution by implementing the rules described above and summarized by the diagram shown in fig. (\ref{figsoftsefi4}), we find the leading order two-loops self energy to be
\bea \Sigma_{lo}(k,\eta,\eta')  & = &    96 \,\lambda^2 ~g^*_\nu(k;\eta)g_\nu(k;\eta')~ (3)~ \Big[g_{\nu}(k,\eta)\,g^*_{\nu}(k,\eta') \Big]~\Big[ \frac{1}{8\pi^2 \Delta\,\eta\,\eta'}\Big]^2+\cdots  \nonumber \\ & = &   \frac{9\,\lambda^2}{2\pi^4} ~\frac{|g^*_\nu(k;\eta)|^2\,|g_\nu(k;\eta')|^2 }{ \Delta^2\,(\eta\eta')^2}+\cdots~ \,.\label{sefi4so}\eea
The dots stand for higher order terms in $\Delta$ that \emph{may} contain logarithmic contributions, furthermore, just as in the case of cubic vertex there is a short distance divergence leading to both mass and wave function renormalization, these would have to be studied in detail and their assessment is beyond the scope of this article, however these are suppressed by  a factor  $\Delta^2\propto \lambda$  because the short distance divergences are the same as in Minkowski space time and of order $\lambda^2$ since these are impervious to the infrared divergences that lead to the poles in $\Delta$.

Again to leading order in $\Delta$ the two loops diagram yields a real contribution to $\Sigma$ therefore no contribution to the mass. The double pole in $\Delta$ confirms the ``infrared counting'' of ref.\cite{holman} but disagrees with the result of the ``plane wave ansatz'' in ref.\cite{rigo} as in the case of the cubic coupling.  Therefore the self-consistency condition becomes $\delta M^2=0$ leading to the same result as the lowest order tadpole (\ref{scfi4},\ref{delfi4}).

\vspace{2mm}

\textbf{Superhorizon modes:} For $-k\eta \ll -k\eta_0 \ll 1$ the $\eta$- integrals yield the same result as for the cubic vertex   using the result (\ref{delfi4})  and to leading order we obtain in this case
\be  C_{1k}(\eta)  \simeq e^{-\gamma(-k\eta)} ~~;~~ \gamma(-k\eta)= \frac{2\lambda }{ \pi^4}~ \Big[\frac{H}{k_{ph}(\eta)}\Big]^{6}\,. \label{C2fi4}\ee Note the power of $\lambda$: the double pole in $\Delta$ leads to the result that the \emph{two loops} contribution is of $\mathcal{O}(\lambda)$ and the terms that have been neglected are at least of $\mathcal{O}(\lambda^{3/2})$. As depicted in fig. (\ref{figsoftsefi4}) the decay process is dominated by the emission and absorption of \emph{two} superhorizon quanta. Thus the power of $\Delta$ in the denominator is directly determined by the number of emitted superhorizon quanta.

Furthermore, in analogy with the interpretation of the width of superhorizon modes for cubic coupling in its relation to the bi-spectrum, it is clear from fig. (\ref{figsoftsefi4}) that the imaginary part of $\Sigma(k,\eta,\eta')$ is related to the trispectrum of scalar fluctuations\cite{nongauss}, and the power $1/k^6$ originates now on a highly squeezed configuration of the \emph{trispectrum}\cite{nongauss} in which two of the momenta $|\vec{p}|, |\vec{q}| < \mu \rightarrow 0$ whereas the other two are $\simeq \vec{k}$. The integral over both bands of wavevectors yields the double pole in $\Delta$, the order of the pole describes the number of superhorizon quanta emitted and absorbed and in this case the number of ``squeezed'' sides of the  trispectrum configuration.

\vspace{2mm}

\textbf{Subhorizon modes:} For $-k\eta_0 \gg -k\eta  \gg 1$ the calculation follows that in the previous case, we find
\be  C_{1k}(\eta)  \simeq e^{-\gamma(-k\eta)} ~~;~~ \gamma(-k\eta)= \frac{9\lambda }{ 8\pi^2}~ \Big[\frac{H}{k_{ph}(\eta)}\Big]^{2}~+\mathcal{O}(\lambda^{3/2})\,. \label{C2fi4sub}\ee As in the case of the cubic coupling we expect that terms of $\mathcal{O}(\lambda^{3/2})$ multiply powers of $k_{ph}(\eta)/H$ which will limit the validity of the leading order term in (\ref{C2fi4sub}) for modes deep inside the Hubble radius. We expect, just as in the previous case that for these deep subhorizon modes, the leading contribution will be determined by the conformal coupling limit, which clearly must be studied in detail for a deeper assessment, a task well beyond the scope of this article, which focuses on the long-wavelength limit.

\section{Conclusions, comments  and further questions}

\subsection{Conclusions:}\label{conclusions}
Motivated by questions on the infrared stability of particle states in de Sitter space time, in this article we    study   massless minimally coupled scalar theories with cubic and quartic interactions   focusing on infrared aspects and the time evolution of single particle states.
 In agreement with previous work\cite{staroyoko,holman,rigo} we find that infrared divergences of massless minimally coupled fields lead to the self-consistent generation of a mass. In the case of the cubic coupling mass generation is a consequence of a radiatively induced expectation value of the scalar field which leads to the formation of a
non-perturbative \emph{condensate} $\langle \phi (\vec{x},\eta) \rangle \propto H\, \big(H/\lambda\big)^{1/3}$ as a result of the infrared divergences. This expectation value induces a  self-consistent mass given by
\be M \propto \lambda^{1/3} \, H^{2/3}  \label{finamaz}\ee   leading   to a de Sitter invariant result
\be \langle \phi^2 \rangle \propto H^2~\Big[\frac{H}{\lambda}\Big]^{2/3}\,. \label{fi2fin3}\ee  For a quartic coupling we find
\be M  \propto \lambda^{1/4}~H ~~;~~\langle \phi^2 \rangle \propto \frac{H^2}{\sqrt{\lambda}} \label{fin2fi4}\ee respectively. The self-consistent mass generation results in that   the infrared divergences in self-energies are now manifest as \emph{poles} in $\Delta = M^2/3H^2$ as a result of the emission and absorption of superhorizon quanta. The two point correlation function of superhorizon fluctuations acquire an \emph{anomalous} dimension $2\Delta$, and $\langle \phi^2 (\vec{x},\eta)\rangle$ is de Sitter invariant.

 The self-consistent treatment of mass generation is combined with the non-perturbative Wigner-Weisskopf method introduced in ref.\cite{boyhol} to extract the time evolution of single particle states. The lack of a global time-like Killing vector entails the lack of kinematic thresholds and  that all single particle states \emph{decay} into quanta of the same field. The decay is dominated by the emission and absorption of superhorizon quanta and hastens when the wave-vector of the single particle state crosses the Hubble radius. The radiatively generated mass regularizes the infrared behavior of the decay rate and leads to a rearrangement of the perturbative expansion \emph{non-analytic in the couplings}.
  We obtain a set of simple rules that yield the leading order contributions to the decay law of single particle states and we find   that the amplitude of single particle states with superhorizon wavelength  decays as $e^{-\gamma(\eta)}~;~\gamma(\eta) \propto (\lambda/H)^{4/3}\,[H/k_{ph}(\eta)]^6~;~\lambda\,[H/k_{ph}(\eta)]^6 $ for cubic and quartic couplings respectively to leading order in $M/H$. The generation of mass as a consequence of the infrared divergences in radiative corrections results in a re-arrangement of the perturbative expansion \emph{non-analytic} in the couplings.

  The power law $1/k^6$ in the decay ``width'' of superhorizon modes is expected on the basis of the relation between the self-energy and the \emph{bi-spectrum or tri-spectrum}\cite{nongauss} of scalar perturbations in the highly squeezed limit for cubic   and quartic coupling  respectively. The order of the pole in $\Delta$ describes the number of superhorizon quanta emitted and absorbed in the intermediate state and also the number of \emph{highly squeezed sides} in the bi-spectrum or tri-spectrum configurations respectively.

 Mass generation and decay entails that all single particle states become \emph{quasiparticles} during de Sitter inflation.


 \subsection{Agreements and disagreements:}

 In the case of quartic coupling $\lambda \phi^4$, our self-consistent one-loop results (\ref{scmassfi4},\ref{fisq})
 agree with those of refs.\cite{holman,serreau} and the one-loop result in\cite{garb} and \emph{qualitatively} with the same power of $\lambda$ with the general results obtained in refs.\cite{raja,garb,staroyoko,riotto} but \emph{not quantitatively}, the disagreement is in numerical factors of order one. We emphasize that the result for the radiatively generated mass (\ref{scmassfi4}) does \emph{not} receive contributions at
 two loops to leading order in $\Delta$ as explained in detail  in section (\ref{subsec:quartic}). The stochastic approach of refs.\cite{staroyoko,riotto} obtains a Fokker-Planck equation for a coarse grained average of the field and extracts $\langle \phi^2 \rangle$ from the asymptotically long time solution of this Fokker-Planck equation, in ref.\cite{staroyoko} such solution features a factor $\Gamma(3/4)/\Gamma(1/4)$   with respect to the one-loop result, a different numerical factor is found in ref.\cite{riotto}. It is not clear   from the stochastic approach which type of diagrams are being resummed by the Fokker-Planck equation, in particular, whereas the two-loop diagram studied in section (\ref{subsec:quartic}) is \emph{non-local} (and retarded), the stochastic approach does not seem to reflect any retardation or non-locality effects. The coarse grained field is treated as a \emph{classical stochastic variable} whereas in a diagrammatic approach the loop integrals reflect the contribution of intermediate states and it is by no means clear (at least to this author) how such contributions contribute to the Fokker-Planck equation (although ref.\cite{richard} provides a field theoretical justification,  precisely how the higher loop retarded contributions are summed up is, again, not clear to this author).   These numerical  factors are also obtained in the result of ref.\cite{raja} which presents a completely different approach: here a Euclidean formulation with compact time on the sphere is implemented,
 the zero mode (on the sphere) is isolated and the path integral of \emph{one single mode} is carried out yielding the
 result $\langle \phi^2 \rangle \propto H^2/\sqrt{\lambda}$ with numerical factors similar to those of the stochastic approach. However, it is by no means clear (at least to this author) how the Euclidean formulation captures any dynamical information which is manifest in the stochastic approach, perhaps the Euclidean compactified description captures the asymptotically long time solution. Furthermore,   in the Euclidean compactified formulation, the zero mode is separated from the other modes by a gap and can be isolated unambiguously, whereas in the Lorentzian version the poles in $\Delta$ arise from the integration of a \emph{band} of infrared wavevectors. Finally, whereas the non-perturbative results beyond one loop obtained in\cite{garb} purport to provide a full resummation of loop diagrams, the final result (eqn. (55) in\cite{garb}) \emph{also disagrees quantitatively} with the results from the stochastic approach {\cite{staroyoko,riotto} and the compactified Euclidean\cite{raja} approach, furthermore, the plane wave ansatze proposed in ref.\cite{garb} is  manifestly a local approximation, and the authors state clearly that they \emph{assume} that the infrared effects can be resummed in a (local) mass term. This is not borne out in the
 results obtained in section (\ref{subsec:quartic}) which reveal that infrared effects also affect the \emph{decay} of quantum states and not only the mass.

 An important aspect that transpires from the two loop result in section (\ref{subsec:quartic}) is that
 the limit $k \rightarrow 0$ is singular, and in fact in separating the integration band $q < \mu \rightarrow 0$ we have manifestly kept the external momentum $k \neq 0$. It is conceivable that the solution to the discrepancy lies in a more thorough treatment of the $k=0$ limit, after all both the stochastic approach and the Euclidean approach consider only the ``zero mode''. But if this is the case there seems to be a discontinuity in the treatment of this mode. Perhaps treating this single mode as a condensate (but without manifestly breaking the underlying symmetry) and separating its contribution as seemingly advocated in ref.\cite{leblond2} is the correct procedure.

 In summary there is general \emph{qualitative agreement} among the various different approaches that in the case of $\lambda \phi^4$ there is a   radiatively generated mass $M \propto \lambda^{1/4} H$ arising from the non-perturbative build-up of infrared effects. However there is a \emph{quantitative disagreement} in the proportionality constant by numerical factors of $\mathcal{O}(1)$. The origin of the discrepancy is difficult to extract because the different approaches that purport to provide a non-perturbative resummation cannot be directly compared, while the stochastic and Euclidean approach share similar numerical factors in the corresponding results, a comparison between the two approaches and with the diagrammatic approach is far from obvious or clear. Understanding the origin of this quantitative discrepancy is certainly worthy of study, but clearly beyond the purview of this work.


\subsection{Comments and further questions:}
\begin{itemize}
\item Powerful results on the asymptotic behavior of correlation functions in Euclidean  de Sitter were obtained in references\cite{donmor,marolf,raja}. In particular for heavy fields with bare mass $M^2 > 9 H^2 /4$ the decay of correlation functions was interpreted as an  imaginary mass, a result confirmed in ref.\cite{leblond}. The continuation from heavy mass to the light mass (or massless) case considered in this article is neither direct nor clear as the superhorizon limit seems to be different, as exhibited in the integral in eqn. (\ref{intesigi}) which is also found in ref.\cite{kroto,leblond}. The analytic continuation from Euclidean results to the Lorentzian de Sitter case if valid also for the massless or light mass case would yield a powerful method to extract the decay of correlators and perhaps establish an equivalence with the decay of quantum states. These aspects clearly merit further and deeper study.

\item A non-perturbative resummation of infrared divergences should yield correlation functions that are well behaved in the superhorizon limit and as $\eta \rightarrow 0^-$. While resummation methods such as the dynamical renormalization group\cite{drg,holman} may ultimately be a successful approach, in this article we followed a different route, namely to study directly the time evolution of states adapting and generalizing  a resummation method that has proven successful in other areas of non-equilibrium phenomena. In ref.\cite{boyhol} the equivalence of this method with the dynamical renormalization group was established in Minkowski space time and it would be fruitful to establish a similar relation in de Sitter cosmology. The main strategy that we advocate can be best described by the example of the two point correlation function of a pion field in Minkowski space time. The pion decays into a lepton pair, the self-energy features a   two lepton threshold and the retarded propagator features a complex pole in the second Riemann sheet as befits a resonance. The time evolution obtained from the frequency Fourier transform   reveals the exponential decay of the correlation function $\propto e^{-\Gamma t/2}$ with $\Gamma$ the decay width. Now consider the pion correlator $\langle \pi(\vec{k},t)\,\pi(-\vec{k},0) \rangle$ introducing a complete set of states between the two fields. Passing to the interaction picture, the one-pion intermediate state is the solution of the Wigner-Weisskopf equation (\ref{CA},\ref{Ckapas}) with a two lepton intermediate state\cite{boyaww},  and decays as $e^{-\Gamma t/2}$ , which is the correct long-time limit from the time Fourier transform of the full propagator. The Wigner-Weisskopf solution is equivalent to a Breit-Wigner approximation to the full propagator\cite{boyhol,boyaww}. We will report on this approach to obtain the correlation functions as generalized to de Sitter space time in a future study.

\item The decay of single particle excitations into superhorizon quanta hastens as the wavevector crosses the Hubble radius. This process entails the \emph{creation} of particles leading to a build-up of the population of the produced particles which  may affect the time evolution of the single particle states as the produced particles \emph{recombine} into the initial state. This possibility requires to study a Boltzmann equation in which both the decay process $1\rightarrow 2$ and its reverse, the recombination  $2 \rightarrow 1$ are taken into account. The question to study is whether  a \emph{detailed balance} emerges where the recombination process balances the decay reaching a steady (or perhaps equilibrium) state. An assessment of these processes requires to obtain a Boltzmann equation as reported in refs.\cite{akmebol,boljap} for the case of heavy fields, but adapted to the massless case and including the self-consistent mass generation mechanism as in the case studied in ref.\cite{boyaboltz} in Minkowski space-time. This program is relegated to future study.

    \item The relation between the decay ``width'' of superhorizon modes and the bispectrum or trispectrum of scalar fluctuations raises an interesting question: are non-gaussianities of curvature perturbations\cite{nongauss} related in any way to the \emph{decay} of either adiabatic or isocurvature superhorizon fluctuations?, if so is there any imprint on the cosmic microwave background anisotropies?. Clearly this question also merits further study.

\end{itemize}

\acknowledgments The author is  supported by NSF grant award   PHY-0852497. He thanks Richard Holman, Andrew Tolley and Matteo Fasiello for illuminating conversations.

\appendix
\section{\label{appsec:SE} Calculation of the Self Energy (\ref{selfa})}

Consider the integral in (\ref{selfa})
\be I [k,\eta,\eta']=  \int_0^\infty \mathcal{H}_{\nu}(p;\eta;\eta')~\Big[\mathcal{F}_{\nu}(k+p;\eta\;\eta')-\mathcal{F}_{\nu}(|k-p|;\eta\;\eta')\Big] p\,dp \label{I} \ee introduce an infrared cutoff $\mu \rightarrow 0$ and write
\be I_\nu[k,\eta,\eta'] \equiv I_0 + I_1-I_2 \label{splits}\ee where
\be I_0 = \int_0^\mu p\,dp \Big[\cdots \Big] \stackrel{\mu \rightarrow 0}{ =}  2 \,k \, \Big[\frac{\Gamma(\nu)}{\pi} \Big]^2\, \Bigg(\frac{4}{\eta\,\eta'}\Bigg)^{\nu} ~\mathcal{H}_{\nu}(k;\eta;\eta') ~~\frac{\mu^{2{\Delta}}}{2\, {\Delta}} \label{I0}\ee and
\bea  I_1 & =  & \int_\mu^\infty \mathcal{H}_{\nu}(p;\eta;\eta')~ \mathcal{F}_{\nu}(k+p;\eta\;\eta')~p\,dp \label{I1}\\ I_2 & = & \int_\mu^\infty \mathcal{H}_{\nu}(p;\eta;\eta')  \mathcal{F}_{\nu}(|k-p|;\eta\;\eta') ~ p\,dp \,.\label{I2}\eea Obviously the total integral $I[k,\eta,\eta']$ is independent of the cutoff $\mu$. Since for $p\rightarrow \infty$
\be \mathcal{H}_{\nu}(p;\eta;\eta') \propto e^{-ip(\eta-\eta')} \label{largep}\ee we introduce a convergence factor
   \be \eta-\eta' \rightarrow \eta-\eta'-i\epsilon ~~;~~ \epsilon \rightarrow 0^+ \,.\label{convergence} \ee

The integral $I_1$ is infrared finite as long as $\mu \neq 0$ and to leading order in $ {\Delta}$ we can set $\nu = 3/2$, namely $ {\Delta} = 0$ therefore
 \bea & & I_1   =    \int_\mu^\infty \mathcal{H}_{\frac{3}{2}}(p;\eta;\eta')~ \mathcal{F}_{\frac{3}{2}}(k+p;\eta\;\eta')~p\,dp  = \nonumber \\ &  &
\frac{4\,e^{-ik(\eta-\eta')}}{\pi^2 \,(\eta\,\eta')^2}\int^\infty_\mu \frac{dp}{p^2}\, e^{-2ip(\eta-\eta')}~\Big[ 1+p^2\,\eta\,\eta'+ip(\eta-\eta')\Big]\Big[\frac{i}{\eta-\eta'-i\epsilon}-\frac{1}{(p+k)\eta\,\eta'} \Big] \label{I1exp}\eea In the integral $I_2$ we must isolate the region $p \simeq k$ extracting the pole in $ {\Delta}$ and outside this region we can replace $\nu = 3/2;  {\Delta}=0$. Therefore we write
\be I_2 = \int_\mu^\infty \Big[\cdots \Big] = \underbrace{\int_\mu^{k-\mu} \Big[\cdots \Big]}_{I^{(a)}_2}+\underbrace{\int_{k-\mu}^k  \Big[\cdots \Big]}_{I^{(b)}_2}+\underbrace{\int_k^{k+\mu} \Big[\cdots \Big]}_{I^{(c)}_2}+\underbrace{\int_{k+\mu}^\infty \Big[\cdots \Big]}_{I^{(d)}_2 }  \label{I2s}\ee In $I^{(a)}_2$ and $I^{(d)}_2$ we can set $\nu=3/2;{\Delta}=0$ since these are infrared finite, namely
\bea I^{(a)}_2  & = &  \int_\mu^{k-\mu}  \mathcal{H}_{\frac{3}{2}}(p;\eta;\eta')~ \mathcal{F}_{\frac{3}{2}}(k-p;\eta\;\eta')~p\,dp \label{I2axp}\\ I^{(d)}_2  & = &  \int_{k+\mu}^{\infty}  \mathcal{H}_{\frac{3}{2}}(p;\eta;\eta')~ \mathcal{F}_{\frac{3}{2}}(p-k;\eta\;\eta')~p\,dp \label{I2dxp} \eea

  For $I^{(b)}_2$ and $I^{(c)}_2$ we find
\be I^{(b)}_2 = I^{(c)}_2 \stackrel{\mu \rightarrow 0}{=} -k\,\frac{\Gamma(\nu)\,\Gamma(\nu-1)}{2\,\pi^2} \Bigg(\frac{4}{\eta\,\eta'}\Bigg)^{\nu} ~\mathcal{H}_{\nu}(k;\eta;\eta') ~~\frac{\mu^{2{\Delta}}}{2\,{\Delta}}\,. \label{I2bc}\ee  therefore keeping the pole in ${\Delta}$ and the leading logarithmic infrared contribution as $\mu\rightarrow 0$ we find
\be I_0 -I^{(b)}_2-I^{(c)}_2 = \frac{4\,k}{\pi} ~ \frac{\mathcal{H}_\frac{3}{2}(k;\eta;\eta')}{\Big(\eta\,\eta'\Big)^\frac{3}{2}}\,
 ~\Bigg[\frac{1}{{\Delta}}+ 2\ln\Big[\frac{\mu}{k}\Big]+ \ln \Big[k^2\,\eta\,\eta'  \Big]+\mathcal{D}[k,\eta,\eta']+\cdots \Bigg] \,,
  \label{IRpart}\ee where
\be \mathcal{D}[k,\eta,\eta'] = -\frac{d}{d\nu} \mathcal{H}_{\nu}[k;\eta;\eta']\Big|_{\nu = \frac{3}{2}}\,. \label{Ddef} \ee
   Furthermore we find
  \be I_1-I^{(a)}_2-I^{(d)}_2 = \frac{4\,e^{-ik(\eta-\eta')}}{\pi^2 \,(\eta\,\eta')^2}~J \label{fins}\ee where
   \bea J  & = &  \int^\infty_\mu \frac{dp}{p^2}\, e^{-2ip(\eta-\eta')}~\Big[ 1+p^2\,\eta\,\eta'+ip(\eta-\eta')\Big]~\Big[\frac{i}{\eta-\eta'-i\epsilon}-\frac{1}{(p+k)\,\eta\,\eta'} \Big] \nonumber \\ & - &  \int^{k-\mu}_\mu \frac{dp}{p^2}~\Big[ 1+p^2\,\eta\,\eta'+ip(\eta-\eta')\Big]~\Big[\frac{i}{\eta-\eta'-i\epsilon}-\frac{1}{(k-p)\,\eta\,\eta'} \Big] \nonumber \\ & - & \int^\infty_\mu  {dp} \, \frac{e^{-2ip(\eta-\eta')}}{(k+p)^2}~\Big[ 1+(k+p)^2\,\eta\,\eta'+i(k+p)(\eta-\eta')\Big]~\Big[\frac{i}{\eta-\eta'-i\epsilon}-\frac{1}{p\,\eta\,\eta'} \Big]\,.\label{bigjota}\eea

    The integrals are straightforward, we find to leading order in ${\Delta}$
   \bea I[k,\eta,\eta'] & = & \frac{4\,k}{\pi} ~ \frac{\mathcal{H}_{\nu}(k;\eta;\eta')}{\big(\eta\,\eta'\big)^{\frac{3}{2}}\,\Delta}~\big[k^2\,\eta\,\eta'\big]^{\Delta}
 \, - \frac{4i\,k}{\pi^2\,\eta\,\eta'}~\frac{e^{-ik(\eta-\eta')}}{(\eta-\eta'-i\epsilon)} \nonumber\\
 &  +  &  \frac{4\,k}{\pi} ~ \frac{\mathcal{H}_\frac{3}{2}(k;\eta;\eta')}{\Big(\eta\,\eta'\Big)^\frac{3}{2}}\, L[k,\eta,\eta'] + \frac{4\,k}{\pi} ~ \frac{\mathcal{H}^*_\frac{3}{2}(k;\eta;\eta')}{\Big(\eta\,\eta'\Big)^\frac{3}{2}} \Bigg[Ci[2k(\eta-\eta')]-i Si[2k(\eta-\eta')]+i\frac{\pi}{2}\Bigg]\nonumber  \\ & + & \frac{8}{\pi^2} \frac{e^{-ik(\eta-\eta')}}{k(\eta\eta')^3} + \mathcal{O}({\Delta}) +\cdots \label{Ifinali}
 \eea where $Ci;Si$ are the cosine and sine integral functions respectively and
 \be L[k,\eta,\eta'] = \ln\Big[2k(\eta-\eta') \Big]  + \gamma + i\frac{\pi}{2} \ee with $\gamma$ the Euler-Mascheroni constant. The infrared logarithm $\ln[\mu]$ cancels out as it can be easily seen from the expressions (\ref{I0}-\ref{I2}) by taking derivatives with respect to $\mu$.

 To write the first term in (\ref{Ifinali}) we have used the definition of $\mathcal{D}[k,\eta,\eta']$ eq. (\ref{Ddef}) and
 \be \frac{4\,k}{\pi} ~ \frac{\mathcal{H}_{\frac{3}{2}}(k;\eta;\eta')}{\big(\eta\,\eta'\big)^{\frac{3}{2}}}\Bigg[\frac{1}{{\Delta}}+ \ln (k^2\,\eta\,\eta' )+\mathcal{D}[k,\eta,\eta']\Bigg] =  \frac{4\,k}{\pi} ~ \frac{\mathcal{H}_{\nu}(k;\eta;\eta')}{\big(\eta\,\eta'\big)^{\frac{3}{2}}\,\Delta}~\big[k^2\,\eta\,\eta'\big]^{\Delta}
 \Big[1+\mathcal{O}(\Delta^2)\Big] \label{reco}\ee

 The second term in the first line in (\ref{Ifinali}) is recognized as the contribution from the massless conformally coupled case\cite{boyhol,boyan}, and  can be written as
\be - \frac{4i\,k}{\pi^2\,\eta\,\eta'}~\frac{e^{-ik(\eta-\eta')}}{(\eta-\eta'-i\epsilon)} = \frac{2ik}{\pi^2 \eta^2}{e^{-ik(\eta-\eta')}}~ \frac{d}{d\eta'} \ln\Big[\big(1-\frac{\eta}{\eta'}\big)+\tilde{\epsilon}^2 \Big]+\frac{4k}{\pi \,\eta^2}~\delta(\eta-\eta') \label{secline} \ee where $\tilde{\epsilon}$ is a physical short distance cutoff that reflects the logarithmic ultraviolet divergence.

\end{document}